\documentclass[useAMS,usenatbib]{mn2e}

\usepackage{natbib, aas_macros}
\usepackage{cite}
\usepackage[dvips]{graphicx}
\usepackage{wrapfig}
\graphicspath{{image/}}
\usepackage{color}
\usepackage{amsmath}
\usepackage{amssymb}
\usepackage{rotating}
\usepackage{appendix}
\bibpunct{(}{)}{;}{a}{}{,} % to follow the A&A style

\newcommand{\HI}{\rm H~{\sc i }}
\newcommand{\HII}{\rm H~{\sc ii }}
\newcommand{\HeI}{\rm He~{\sc i }}
\newcommand{\HeII}{\rm He~{\sc ii }}
\newcommand{\HeIII}{\rm He~{\sc iii }}
\newcommand{\TB}{\delta T_{\rm b}}
\newcommand{\MSUN}{\rm M_{\odot}}
\newcommand{\XHI}{x_{\rm HI}}
\newcommand{\XHII}{x_{\rm HII}}

\newcommand{\TS}{T_{\rm S}}
\newcommand{\TK}{T_{\rm K}}
\newcommand{\TCMB}{T_{\rm CMB}}
\newcommand{\lya}{\rm {Ly{\alpha}}}
\newcommand{\OmegaB}{\Omega_{\rm B}}
\newcommand{\Omegam}{\Omega_{\rm m}}

\title[21 cm signal from cosmic dawn]{21 cm signal from cosmic dawn: imprints of spin temperature fluctuations and peculiar velocities}

\author[Ghara, Choudhury \& Datta]{Raghunath Ghara$^1$\thanks{Email: raghunath@ncra.tifr.res.in}, T. Roy Choudhury$^1$\thanks{Email: tirth@ncra.tifr.res.in} and Kanan K. Datta$^{1,2}$\thanks{Email: kanan@ncra.tifr.res.in}\\
$^1$ National Centre for Radio Astrophysics, TIFR, Post Bag 3, Ganeshkhind, Pune 411007, India\\ 
$^2$ Department of Physics, Presidency University, 86/1 College Street, Kolkata - 700073, India } 

\begin{document}

\date{Accepted ?; Received ??; in original form ???}

\pagerange{\pageref{firstpage}--\pageref{lastpage}} \pubyear{?}

\maketitle

\label{firstpage}

% Abstract------------------------------------------------------------

\begin{abstract}

The 21 cm brightness temperature $\TB$ fluctuations from reionization promise to provide information on the physical
processes during that epoch. We present a formalism for generating the $\TB$ distribution using dark matter
simulations and an one-dimensional radiative transfer code. Our analysis is able to account for the spin temperature $\TS$
fluctuations arising from inhomogeneous X-ray heating and $\lya$ coupling during cosmic dawn. The $\TB$ power
spectrum amplitude at large scales ($k \sim 0.1$ Mpc$^{-1}$) is maximum when $\sim 10\%$ of the gas (by volume) is heated 
above the cosmic microwave background temperature. The power spectrum shows a ``bump''-like feature during cosmic dawn and its location measures the 
typical sizes of heated regions. We find that the effect of peculiar velocities on the power spectrum is negligible at large 
scales for most part of the reionization history. During early stages (when the volume averaged ionization fraction
$\lesssim 0.2$) this is because the signal is dominated by fluctuations in $\TS$. For reionization models that are solely
driven by stars within high mass ($\gtrsim 10^9\, \MSUN$) haloes, the peculiar velocity effects are prominent only at
smaller scales ($k \gtrsim 0.4$ Mpc$^{-1}$) where patchiness in the neutral hydrogen density dominates the signal. The
conclusions are unaffected by changes in the amplitude or steepness in the X-ray spectra of the sources.

\end{abstract}

\begin{keywords}
intergalactic medium - radiative transfer - cosmology: theory - cosmology: dark ages, reionization, first stars - galaxies: formation -  X-rays: galaxies

\end{keywords}

\section{Introduction}
\label{intro}

The birth of first stars, galaxies, quasars is one of the landmark events in the history of the Universe. It is believed that radiation produced by these sources spread through the intergalactic medium (IGM) and changed its thermal and ionization state  completely. The onset of first sources of light which marked the end of the `dark ages' and changed the IGM thermal state is often termed as `cosmic dawn'. Subsequent period when the neutral hydrogen ( \HI) was ionized is popularly known as the epoch of reionization (EoR). 

Unfortunately we have a very little knowledge about this event. Many important questions such as the exact timing, thermal and ionization state of the IGM, properties of the first sources, sinks, feedback effect, impact on the structure formation of the Universe etc. are largely unknown. This is because the signal coming from this epoch is so faint that even the most sensitive instrument existing could only provide limited information of the event. Nevertheless, observations of redshift $z \gtrsim 6$ quasars absorption spectra \citep{Gunn65, becker01, fan03, fan2006, Goto11} and the cosmic microwave background radiation (CMB) \citep{Komatsu11, Planck2013} suggest that the event probably took place around $6 < z < 15$ \citep{Choudhury06a, fan2006, Malhotra06, mitra2011, mitra2012}.

Observations of the redshifted \HI 21-cm signal from the IGM is a powerful probe and  is believed to provide us with enormous amount of information about the epoch \citep{Furlanetto2006, Morales10, Pritchard12}. A huge effort is in place to measure the redshifted 21 cm signal from the cosmic dawn and EoR. First generation of low frequency radio telescopes like Low Frequency Array (LOFAR)\footnote{http://www.lofar.org/} \citep{van13}, Murchison Widefield Array (MWA)\footnote{http://www.mwatelescope.org/} \citep{bowman13, tingay13}, Giant Metrewave Radio Telescope (GMRT) \footnote{http://www.gmrt.tifr.res.in}\citep{ghosh12, paciga13}, Precision Array for Probing the Epoch of Reionization (PAPER)\footnote{http://eor.berkeley.edu/} \citep{parsons13}, 21 Centimetre Array (21CMA)\footnote{http://21cma.bao.ac.cn/} have started observing with the aim to detect the signal from the EoR. Due to lack of lower frequency band these instruments might just miss the era of cosmic dawn when the very first sources were formed and changed the thermal state of the IGM. The extremely sensitive next generation telescope, the Square Kilometre Array (SKA)\footnote{http://www.skatelescope.org/}, is expected to measure the signal at even lower frequencies which, in addition to the EoR, should also be able to probe the cosmic dawn \citep{mellema13}. 

The \HI 21 cm signal will be observed against the CMB radiation. The signal will be observed in either emission or absorption depending on whether the \HI spin temperature is higher or lower than the background CMB temperature. It is often assumed that the spin temperature is highly coupled with the IGM kinetic temperature and much higher compared to the CMB temperature right from the birth of first light sources i.e, when reionization process starts \citep{furlanetto04, McQuinn2005, mesinger07, choudhury09, datta2012a, battaglia13, iliev14}. This leads to a situation in which the 21 cm signal is independent of the exact value of the spin temperature. However, these assumptions may not hold during the initial stages of reionization or the cosmic dawn. Although a small amount of $\lya$ photons is enough for establishing the coupling, it is possible that even that small number of $\lya$ photons is not available at every location in the IGM. The heating of the IGM is also highly dependent on the nature, total number and spectra  of X-ray sources at cosmic dawn which are all unknown. In a situation where the collisions (between \HI atoms or \HI atoms and free electrons) become inefficient and $\lya$ photons are only confined near the sources, the coupling between the IGM kinetic temperature and \HI spin temperature becomes inhomogeneous. Even if the coupling is strong and complete, the heating could be incomplete and inhomogeneous due to lack of X-ray photons. This would lead to fluctuations in the spin temperature and consequently additional features in the 21 cm signal. Recently, there have been several attempts in order to understand the effect of spin temperature fluctuations  on the observed \HI 21 cm signal during the cosmic dawn \citep{barkana05b, Chuzhoy07, semelin07, santos08, baek09, thomas11, McQuinn2012}.  Models with complete $\lya$ coupling and heating predicts  the variance of \HI brightness temperature fluctuations to be a few 10 ${\rm mK}^2$, while the same quantity for inhomogeneous $\lya$ coupling and heating model could exceed few 100 ${\rm mK}^2$.

Efforts are also in place to understand the heating from different kind of X-ray sources such a miniquasars \citep{thomas11}, X-ray binaries \citep{Fialkov14,ahn2014}, and thermal emission from hot interstellar medium \citep{Pacucci2014}.  Because of their large mean free path, the hard X-ray photons can penetrate a few tens of Mpcs in the IGM \citep{shull1985}, while soft X-ray photons will be absorbed within a smaller distance from the sources and as a result the heating will be very patchy. The heating pattern will be very different in the case of high mass X-ray binaries (HMXBs) when compared to miniquasars or hot interstellar medium as they do not contain very large amount of soft X-rays \citep{Fialkov14, Pacucci2014}. It has also proposed that measurements of \HI 21 cm power spectrum from the cosmic dawn will reveal the amount of X-ray background \citep{christian13} and the spectral energy distribution (SED) of X-ray sources \citep{Pacucci2014}.

In this paper, we study  the effects of inhomogeneous $\lya$ coupling and IGM heating on the \HI 21 cm signal from the reionization epoch and cosmic dawn using a semi-numerical code which is primarily based on the algorithm presented in \citet{thomas08} and \citet{Thom09}.  We use slightly different methods for the $\lya$ coupling and heating calculation. We consider miniquasar like objects ( defined as galaxies with intermediate mass black holes (BHs)  in the range of $10^3-10^6\, \MSUN$) as X-ray heating sources and focus on statistical quantities such as the variance and power spectrum of the \HI brightness temperature fluctuations which are among the primary goals of instruments like LOFAR and SKA.  

Our major effort, in this paper, has gone into understanding the effect of the peculiar velocity on the \HI 21-cm signal from the cosmic dawn.  The peculiar velocity has a significant impact on the reionization and pre-reionization \HI 21-cm signal \citep{bharadwaj04}. This also makes 21-cm power spectrum anisotropic \citep{barkana05a, Majumdar13}.  Recently, it has been shown that the peculiar velocity can boost the \HI power spectrum by a factor of $\sim 5$ at large scales during the initial stages of reionization when $\XHII \lesssim 0.2$ \citep{mao12, Jensen13}. During the same period power spectrum becomes highly anisotropic which is detectable with LOFAR 2000 h of observations. It was also suggested that such observations could tell us whether reionization occurred inside-out or outside-in \citep{Jensen13, Majumdar13}. However, all results described above are based on the assumption that the spin temperature is much higher than the CMB temperature. Here we investigate how the above results change once the heating and $\lya$ coupling is calculated self consistently. As we see later, implementing the peculiar velocity effect on the signal during this epoch is slightly different compared to the case when the spin temperature is much higher than the CMB temperature. Apart from that, we investigate how  peaks in the power spectrum can be used to extract information about the ionization state and size of the `heated bubbles'. 

The plan of this paper is as follows: In section $\ref{21cm}$ we have briefly described the properties of the redshifted 21 cm signal from \HI. We then discuss the detailed methodology for modelling the signal, including the $N$-body simulation used for the present study (section \ref{numsim}), the source model (section \ref{source}), radiative transfer code for generating the maps around isolated sources (section \ref{rt}), the generation of global maps (section \ref{global_map}) and modelling the redshift space distortion (section \ref{rsd}). In section \ref{res} we discuss the main results of our analyses.  Results related to reionization models containing only high mass sources have been described in section \ref{large_halo}. The globally averaged ionization and heating properties of our models are discussed in section \ref{res:global_hist} while the fluctuations in the 21 cm signal are described in section \ref{res:tb}. The main results of our work, i.e.,  the effects of the peculiar velocities on the 21 cm signal are discussed in section \ref{res:rsd}. In section \ref{res:X_rsd}, we see whether our conclusions are unchanged when the X-ray properties of the sources are varied.  In section \ref{small_halo}, we have described the effect of small mass haloes on the results,  and we have checked the robustness of our results with respect to the resolution of the simulation box in section \ref{small_box}. We summarize and discuss our main results in section \ref{conc}. Throughout the paper, we have used the cosmological parameters $\Omegam=0.32$, $\Omega_\Lambda=0.68$, $\OmegaB=0.049$, $h=0.67$, $n_{\rm s}=0.96$, and $\sigma_8=0.83$ which are consistent with the recent results of $Planck$ mission \citep{Planck2013}.

\section{Simulating the 21 cm signal}
\label{21cm}

Usually the redshifted 21 cm signal from neutral hydrogen is measured in terms of the deviation of 21 cm brightness temperature from the brightness temperature of background CMB radiation along a line of sight. The differential brightness temperature observed at a frequency
$\nu_{\rm obs}$ along a direction $\mathbf{\hat{n}}$ is given by \citep{madau1997,Furlanetto2006}
\begin{eqnarray}
\TB (\nu_{\rm obs}, \mathbf{\hat{n}}) \!\!\!\! &\equiv& \!\!\!\! \TB (\mathbf{x}) = 27 ~ \XHI (z,\mathbf{x}) [1+\delta_{\rm B}(z,\mathbf{x})] \left(\frac{\OmegaB h^2}{0.023}\right) \nonumber\\
&\times& \!\!\!\!\left(\frac{0.15}{\Omegam h^2}\frac{1+z}{10}\right)^{1/2}\left[1-\frac{\TCMB(z)}{T_{\rm S}(z,\mathbf{x})}\right]\,\rm{mK},
\nonumber \\
\label{brightnessT}
\end{eqnarray}
where $\mathbf{x} = r_z \mathbf{\hat{n}}$ and $1+z = 1420~{\rm MHz}/\nu_{\rm obs}$, with $r_z$ being the comoving radial distance to redshift $z$.
The quantities $\XHI(z,\mathbf{x})$ and $\delta_{\rm B}(z,\mathbf{x})$ denote the neutral hydrogen fraction and the density contrast in baryons respectively at point $\mathbf{x}$ at a redshift $z$. The CMB temperature at a redshift $z$ is denoted by $\TCMB(z)$ = 2.73 $\times (1+z)$ K and $\TS$ is the spin temperature of neutral hydrogen.
We have omitted the effect of line of sight peculiar velocities \citep{bharadwaj04, barkana05a} in the above expression which 
essentially maps the point $\mathbf{x}$ in real space to a point $\mathbf{s}$ in the redshift-space, the mapping being determined by the line of sight peculiar velocity field. 
We will discuss how to account for this effect in Section \ref{rsd}.

The primary goal of the first generation radio telescopes is to measure the fluctuations in $\TB$ using the spherically averaged power spectrum $P(k)$ which is defined as
\begin{equation}
\langle \hat{\TB}(\mathrm{\bf k}) \hat{\TB}^{\star}(\mathbf{k'})\rangle = (2 \pi)^3 \delta_D(\mathbf{k - k'}) P(k),
\label{ps}
\end{equation}
where $\hat{\TB}(\mathrm{\bf k})$ is the Fourier transform of $\TB(\mathbf{x})$ defined in equation (\ref{brightnessT}). The dimensionless power spectrum is defined as $\Delta^2(k)=k^3P(k)/2\pi^2$ which also represents the power per unit logarithmic interval in $k$.

The spin temperature $\TS$ is determined by the coupling of neutral hydrogen gas with CMB photons by Thomson scattering, $\lya$ coupling and collisional coupling. Considering all these coupling effects, the spin temperature can be written in the following form \citep{field58,Furlanetto06a}
\begin{equation}
\TS^{-1}=\frac{\TCMB^{-1}+x_\alpha T_\alpha^{-1}+x_c \TK^{-1}}{1+x_\alpha+x_c},
\label{spinT}
\end{equation}
where $\TK$ is the kinetic  temperature of the gas and $T_{\alpha}$ is the colour temperature of the $\lya$ photons which, in most cases of interest, is coupled to $\TK$ by recoil during repeated scattering. The quantities $x_c$ and $x_{\alpha}$ are the coupling coefficients due to collisions and $\lya$ scattering, respectively. 
The collision efficiency $x_c$ includes both the collisions between neutral hydrogen atoms (${\rm H-H}$) and hydrogen atom with free electrons in the medium (${\rm H-e}$) and can be written as \citep{hirata2006col},
\begin{equation}
x_c=\frac{4T_\star}{3A_{10}\TCMB}\left[\kappa^{\rm HH}(\TK)n_{\rm H} +\kappa^{\rm eH}(\TK)n_e\right],
\label{xc}
\end{equation}
where $A_{10} = 2.85 \times 10^{-15} s^{-1}$ is the spontaneous Einstein A-coefficient, $T_\star = h\nu_{\rm 21cm}/k = 0.0681 K$, $n_{\rm H}$ and $n_e$ be the local number densities of neutral hydrogen and electrons respectively. The table of $\kappa^{\rm eH}$ as a function of $\TK$ is taken from \citet{furlanettoes2006} and table of $\kappa^{\rm HH}$ is taken from \citet{zygelman2005, allison1969}.  

The most important process which couples $\TS$ to $\TK$ during reionization is the Wouthysen-Field effect \citep{wouth52, field58, madau1997,  hirata2006lya, chuzhoy2006}. The $\lya$ coupling coefficient in this case is given by
\begin{equation}
x_\alpha=\frac{16\pi^2T_\star e^2 f_\alpha}{27A_{10}\TCMB m_e c} J_\alpha,
\label{xalpha}
\end{equation}
where $J_\alpha$ is the $\lya$ flux density, $f_\alpha=0.4162$ is the oscillator strength for the $\lya$ transition,  $m_e$ and $e$ are the mass and charge of the electron respectively. In many studies of reionization, particularly those dealing with later stages  \citep{furlanetto04, McQuinn2005, mesinger07, choudhury09, datta2012a, battaglia13, iliev14}, one assumes the $\lya$ coupling to be highly efficient and uses the approximation $\TS \approx \TK$. In addition, if the IGM is assumed to be heated substantially compared to the CMB ($\TK \gg \TCMB$), then the term $(\TS - \TCMB)/\TS \to 1$ in equation (\ref{brightnessT}) and hence one obtains the simple expression where $\TB$ tracks the neutral hydrogen distribution $\XHI (1 + \delta_{\rm B})$. These assumptions, however, have been shown not to hold in early stages of reionization where the $\lya$ coupling may not be uniformly strong in all locations and not all regions of the IGM will be heated uniformly \citep{santos08, Baek2010}. This would lead to variations in $\TS$ and hence would affect the fluctuations in the 21 cm signal. We take all these effects into account in this paper using a combination of $N$-body simulations and an one-dimensional radiative transfer code.

%\\\\\\\\\\\\\\\\\\\\\\\\\\\\\\\\\\\\\\\\\\\\\\\\\\\\\\\\\\\\\\\\\\\\\\\\\\\\\\\\\\\\\\\\\\\\\\\\\\\\\\\\\\\\\\\\\\\\\\\\\\\\\\\\\\\\\\\\\\\\\\\\\\\\\\\\\\\\\\\\\\

\subsection{Numerical simulations} 
\label{numsim}

The method we have used for simulating the 21 cm brightness temperature signal is essentially based on (i) obtaining the dark matter density field and the distribution of collapsed haloes from a $N$-body simulation, (ii) assigning luminosities to these dark matter haloes and (iii) using a one-dimensional radiative transfer code to obtain the neutral hydrogen and spin temperature maps.

We have performed dark matter $N$-body simulations using the publicly available {\sc cubep}$^3${\sc m}\footnote{\tt http://wiki.cita.utoronto.ca/mediawiki/index.php/CubePM} \citep{Harnois12} which is essentially a massively parallel particle-particle-particle-mesh (P$^3$M) code. Initialization of the particle positions and velocities at redshift $z=200$ was done by using {\sc camb} transfer function\footnote{\tt http://camb.info/}  \citep{lewis00}  and employing Zel'dovich approximation. The fiducial simulation used in this work contains $768^3$ particles in a box of size 100 $h^{-1}$ cMpc with $1536^3$ grid points. The mass resolution of the dark matter particles in the simulation is $1.945\times 10^8\, h^{-1}\, \MSUN$. The simulation generates snapshots at $25 \geq z \geq 6$ in equal time gap of $10^7$ years. The output at each snapshot consists of the density and velocity fields in a grid which is 8 times coarser than the simulation grid. The code is also equipped with a run time halo finder which identifies haloes within the simulation volume using spherical overdensity algorithm. We assume that the smallest halo at least contains 20 dark matter particles. As far as the baryonic density field is concerned, we simply assume that the baryons trace the dark matter, i.e., each dark matter particle is accompanied by a baryonic particle of mass  $\left(\OmegaB/\Omegam \right) \times M_{\rm part}$. Though this assumption is not valid in very small scales, i.e., scales comparable to or smaller than the local jeans scale, it probably works fine at large scales which are of our interest.

 The main difficulty with our simulation is that the volume does not contain small mass haloes (e.g., those with $M_{\rm halo} \sim 10^8\, \MSUN$) which are believed to be driving the reionization at early stages. Resolving such small haloes requires simulations of very high dynamic range which are beyond the computing power we have access to. In order to address this issue, we have used a sub-grid model to include the small mass haloes in our simulation box, as described in appendix \ref{appendix}. The minimum halo mass resolution achieved in this case is $\sim 10^8\, \MSUN$, which is helpful in probing the effects of small mass haloes at early stages.  We first present and discuss our results using the fiducial box containing only large mass sources, and then discuss whether missing out the small haloes makes any difference to our conclusions in a separate section (Section \ref{small_halo}). 

%........................................

\subsection{Source selection}
\label{source}
The stars residing in the galaxies are believed to be the major source of ionizing photons that complete the hydrogen reionization process of the universe. The dark matter haloes are the most suitable place to form galaxies but not all of them will contain luminous sources. For galaxy formation to proceed, the gas is required to be cooled below their virial temperature by atomic (or molecular) cooling which may not be possible in the smallest mass haloes. Typically, the minimum mass of haloes which can cool via atomic hydrogen is $\sim 10^8\, \MSUN$ while the same value can be much smaller $\sim 10^6\, \MSUN$ in the presence of hydrogen molecules. In addition, there could be further suppression of star formation in haloes lighter than $\sim 10^9\, \MSUN$ which are residing in ionized regions because of radiative feedback. Since the smallest halo in our fiducial simulation has mass $3.89 \times 10^9\, h^{-1}\, \MSUN$, we are unable to tackle all these complications self-consistently, hence we assume all haloes in our simulation box to form stars at all redshifts. We will later (Section \ref{small_box}) discuss the simulation where haloes having masses as small as $\sim 10^8\, h^{-1}\, \MSUN$ are resolved

The relation between the dark matter halo mass and the galaxy luminosity of these early galaxies are all very uncertain. The fraction of baryons $f_\star$ residing within the stars in a galaxy depends on the metallicity and mass of the galaxy. There is no well known relation between the stellar mass and total mass of the galaxies at very high redshifts. For simplicity, here  we have assumed $f_\star$ to be constant throughout the reionization epoch and its value is chosen such that the resulting reionization history is consistent with the constraints obtained from CMB polarization measurements. The stellar mass of a galaxy corresponding to a dark matter halo of mass $M_{\rm halo}$ is
\begin{equation}
M_\star=f_\star \left(\frac{\OmegaB}{\Omegam}\right) M_{\rm halo}.
\end{equation}

Given $M_\star$, one can calculate the spectral energy distribution (SED) of stellar sources in a galaxy using stellar population synthesis codes. However, the SED will depend upon the initial metallicity and the stellar IMF, both of which evolve with time. For example,
stars in very first galaxies are expected to be metal poor \citep{Lai07,Finkelstein09b} and short-lived \citep{Meyn05}. Eventually they enrich the ISM with metals which changes the nature of subsequent star formation. Since tracking the evolution of metallicity self-consistently is not straightforward, we have taken the best fit mass metallicity relation from \citet{dayal09a} and \citet{dayal10}
\begin{equation}
\frac{Z}{Z_\odot}=(0.25-0.05 \Delta z) ~\log_{10}\left(\frac{M_\star}{\MSUN}\right)-(2.0-0.3\Delta z),
\label{met}
\end{equation}
where $\Delta z = (z-5.7)$. 
The evolution of stellar IMF of high redshift galaxies too is not well constrained. For our study we assume that the stars follow a Salpeter IMF with mass range 1 to 100 $\MSUN$. 

The UV and NIR spectral energy distributions  of the stellar sources in galaxies are generated using the code {\sc pegase2}\footnote{\tt http://www2.iap.fr/pegase/} \citep{Fioc97} which computes the galactic SED using standard star formation scenarios for different initial metallicities, IMF and star formation history at different epochs. The lifetime of the stars in the galaxy may vary with metallicity and mass \citep{Meyn05}.  For convenience, we set the  stellar lifetime to be $10^7$ years which is the time difference between two simulation snapshots. 

\begin{figure}
\begin{center}
\includegraphics[scale=0.7]{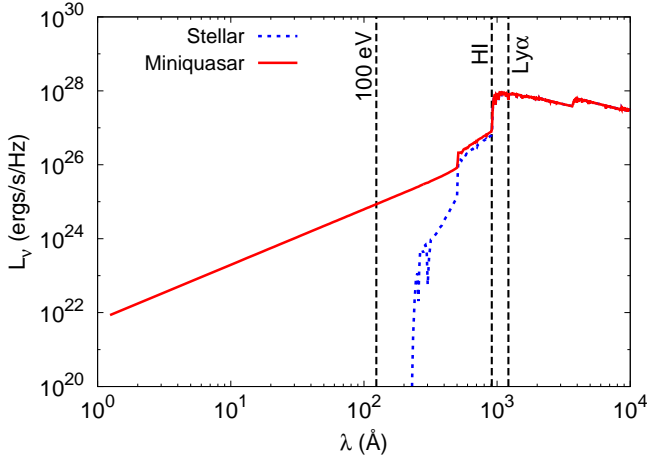}
    \caption{The SED of a galaxy of mass $10^8\, \MSUN $ and metallicity $0.1$ $Z\odot$. The dotted blue curve shows the stellar-like component, while the solid red curve shows the SED of the miniquasar like component. We have assumed the miniquasar to emit 5$\%$ of the UV energy in X-ray band with a power law SED (in this case the power law index $\alpha$ is 1.5).  The vertical dashed lines correspond to energies 10.2 eV, 13.6 eV and 100 eV respectively.}
   \label{spec}
\end{center}
\end{figure}

The dotted blue curve in Figure \ref{spec} shows the intrinsic SED of the stellar component in a galaxy with stellar mass $10^8\, \MSUN $ with metallicity 0.1 $Z_\odot$. The SED peaks around hydrogen ionization wavelength and falls sharply for higher energies ($\gtrsim$ 50 eV). Thus, in the absence of any other processes, these galaxies can only ionize \HI and \HeI, however they are unlikely to be efficient sources of \HeII ionization and heating of the IGM.

These galaxies can produce photons of higher energies if, e.g., they harbour accreting (supermassive or intermediate mass) black holes at the centre. In that case, these galaxies will behave as miniquasars and will produce X-rays with power-law SED $E^{-\alpha}$ \citep{elvis94, laor97, vanden01, vignali03}. There could be other sources of X-rays in galaxies like the HMXBs. These have a very different SED in the X-ray band \citep{frag1, frag2} and thus show very different effects on heating and ionization \citep{Fialkov14}. Another major source of X-ray photons could be the hot interstellar medium within early galaxies \citep{Pacucci2014}. In this work, we concentrate only on the effect of miniquasar-like sources and leave the studies of other sources to a future work\footnote{It turns out that the impact of X-rays from the hot interstellar medium is similar to those from the miniquasar like sources, as we will see later in this paper. This is because of the fact that the hot interstellar medium produces significant number of soft X-ray photons similar to miniquasar like sources. The HMXBs, on the other hand, show very different signatures since they do not contain any significant amount of soft X-rays.}.

Let us assume the UV spectrum to be spanning from 10.2 eV to 100 eV in energy band, and the X-ray spectrum to span from 100 eV to 10 keV. Let $f_X$ be the ratio of X-ray to UV luminosity from the sources\footnote{ The parameter $f_X$ in our model can be related to the black hole (BH) to galaxy mass ratio. Recent observations of high redshift quasars show that accretion rate of black holes are close to the Eddington limit (e.g., \citealt{Willott10b}). In addition observations of local galaxies show that the mass ratio of the BH and galaxy is $\sim 10^{-3}$ (e.g., \citealt{Rix04}). If we fix $\alpha = 1.5$ in our model, we find that for a metal free source the BH to galaxy mass ratio is $\sim 2 \times 10^{-3}$ for $f_X = 0.1$. Thus we choose $f_X = 0.1$ as a upper bound for our study.}
\begin{equation}
f_X = \frac{\int\limits_{100~{\rm eV}}^{10~{\rm keV}} I(E) dE}{\int\limits_{10.2~{\rm eV}}^{100~{\rm eV}} I(E) dE},
\end{equation}
where $I(E)$ is the SED of the galaxy. Let us denote the SED of the stellar component by $I_\star(E)$, which in our case is computed using the stellar population synthesis code {\sc pegase2}. We have seen from Figure \ref{spec} that this component does not contribute significantly to the X-rays, i.e., to the numerator of the above equation. The SED of the miniquasar-like component is assumed to  have a power law form
\begin{equation}
I_q(E) = A ~ E^{-\alpha}, 
\end{equation}
where $A$ is a normalization constant to be determined in terms of $f_X$ and $\alpha$. Let $L_\star$ be the luminosity of the stellar component in the UV band, i.e., $L_\star = \int_{10.2~{\rm eV}}^{100~{\rm eV}} I_\star(E)~{\rm d} E$. Then, it is straightforward to show that the constant $A$ is determined by the relation
\begin{equation} 
A = \frac{f_X L_{\star}}{\left ( \int\limits_{100~{\rm eV}}^{10~{\rm keV}} E^{-\alpha} dE - f_X \int\limits_{10.2~{\rm eV}}^{100~{\rm eV}} E^{-\alpha} dE \right )}.
\end{equation}
We have varied the X-ray properties (i.e., the parameters $f_X$ and $\alpha$) of the sources to study the effects on various quantities of interest. 
However, while choosing the values for $f_X$ and $\alpha$ it should be kept in mind that the normalization coefficient $A$ must be positive.

%..............................................

\subsection{Radiative transfer around an isolated source}
\label{rt}

As the first sources of light appear in the universe, they start to ionize the surrounding IGM and create ionized bubbles around them. In addition, the UV and X-ray radiation from these sources would heat up the medium. In order to simulate the ionization and heating patterns around ionizing sources, we have developed a code which closely follows the treatment of \citet{thomas08, thomas11}. Essentially, the global ionization and temperature distributions are simulated in two steps: in the first step, we generate the ionization and temperature patterns as a function of time and distance around isolated sources surrounded by a uniform density field, and in the second step, we account for possible overlaps between these individual patterns around all the sources in the simulation box. We refer the readers to the original papers of \citet{thomas08, thomas11} for the details of the algorithm. Here we highlight some of the important features of the method and the modifications we have made to the method.

\begin{itemize}

\item Let us consider a galaxy of stellar mass $M_\star$ surrounded by a uniform IGM. Following the procedures outlined in the previous section, we can estimate the photons production rate at every energy band of interest. We then assume that the ionization and heating pattern around this isolated galaxy would be spherical in nature, and hence a one-dimensional radiative transfer setup is sufficient to solve the problem. The motivation for this assumption comes from the fact that at early stages of reionization the number density of the sources is low and the bubbles are believed to be separated from each other. Radiative transfer simulations studying the impact of the sources on the surrounding IGM indicate that the ionized bubbles around galaxies were probably almost spherical in nature before the overlap started \citep{kuhlen2005, thomas08, Alvarez10}. 

\item The number densities of \HI and \HII along with those of \HeI, \HeII and \HeIII are calculated as a function of distance $d$ from the source by solving the relevant rate equations. The most relevant processes in calculating the fraction of different species are photoionization by UV radiation, secondary ionization arising from high-energy free electrons and recombination. 

\item These rate equations are supplemented by the evolution equation for $\TK$.  The initial temperature is assumed to be uniform with a value $T^{i}_{\rm K}(z_i) = 2.73~{\rm K} \times (1+z_i)^{2}/(1+z_{\rm dec})$, i.e., we assume that the gas temperature was coupled to the CMB temperature till redshift $z_{\rm dec}$ and then evolved as $(1+z)^2$ due to adiabatic expansion of the gas. In this study, we have fixed $z_{\rm dec}$ to be 150 \citep[e.g.,][]{Furlanetto2006}. The most prominent heating processes are the photoheating due to UV and X-ray flux. The UV-heating is most prominent near the boundary of the ionized and neutral regions, while X-ray heating is substantial within the partially ionized and neutral regions because of large mean free path of high energy photons. In addition, X-rays can increase the kinetic temperature by sufficient Compton scattering with the free electrons in the medium. Among the various cooling processes described in \citet{thomas08}, the main cooling of the IGM comes from the expansion of the universe.

\item The UV and X-ray radiation flux from the central source is calculated self-consistently accounting for the $d^{-2}$ dilution and also the effects of optical depth along the photon path.

\item In addition to the ionizing flux, we also keep track of the $\lya$ flux $J_{\alpha}$ around the source. The  $\lya$ flux is assumed to decrease as $d^{-2}$ from the source. In addition to the continuum $\lya$ photons from the stellar sources, another source of $\lya$ photons is the X-rays from the source. In this case a fraction of energy from the primary electrons is spent in exciting the \HI, which then generate $\lya$ photons on relaxation. Since the $\lya$ photons travel large distances without being absorbed, we account for the redshift of photons having energies larger than the $\lya$ frequency produced in the source. We have also restricted the $\lya$ photons to propagate with speed of light for a time equal to the age of the hosting halo.

\item We store profiles of the fraction of different species, the kinetic temperature and the $\lya$ flux for a wide range of galaxy stellar masses, redshifts, density contrast, X-ray to UV luminosity ratio and X-ray SED spectral index. These catalogues are to be used later while constructing global maps of the 21 cm brightness temperature.

\end{itemize}

\begin{figure}
\begin{center}
\includegraphics[scale=0.52]{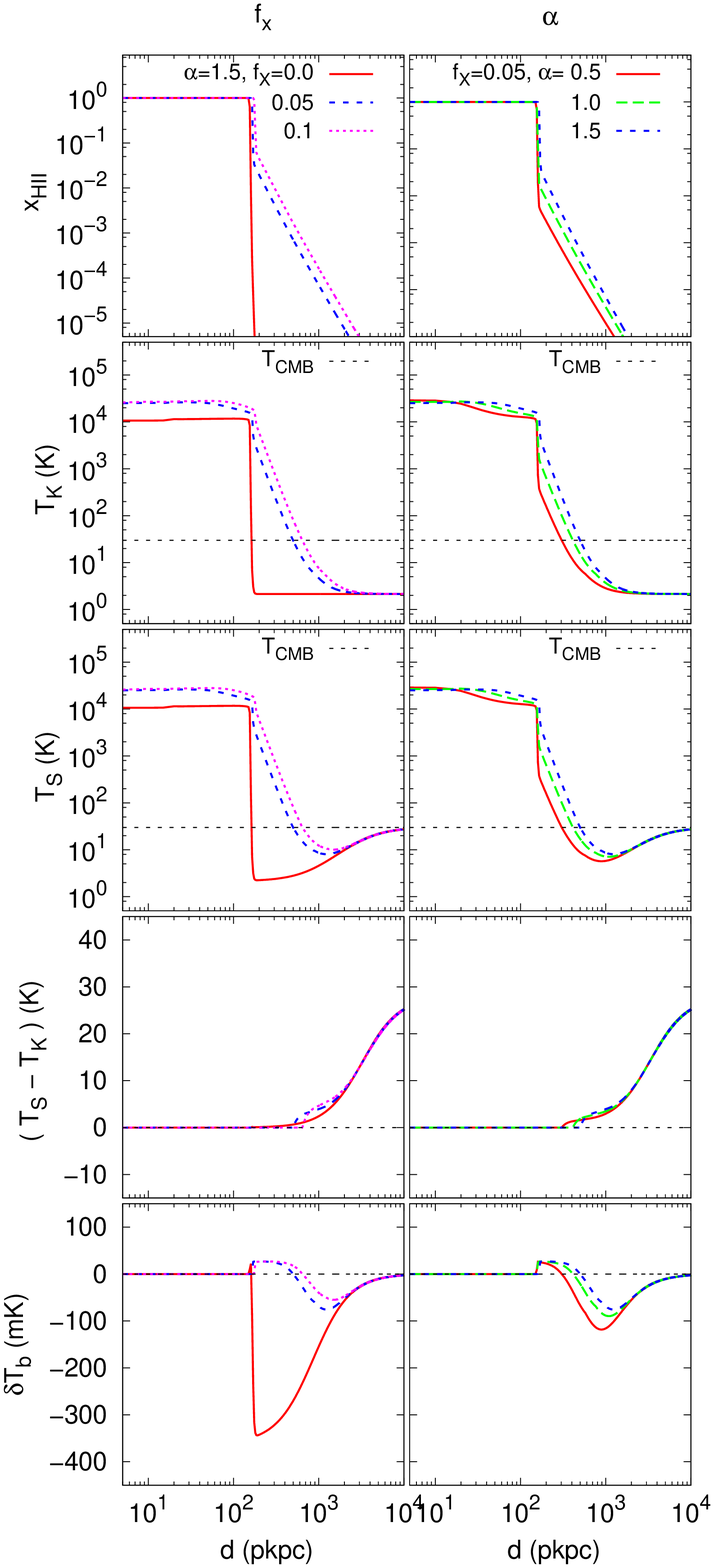}
    \caption{The top to bottom rows show the ionization fraction ($\XHII$), kinetic temperature ($\TK$), spin temperature ($\TS$), the difference between spin and kinetic temperatures ($\TS - \TK$) and differential brightness temperature ($\TB$)  distribution around a galaxy with stellar mass $10^8\, \MSUN$ at a time 10 Myr after the source starts radiating. The surrounding IGM is taken to be uniform with mean density of the universe at redshift 10. In first column, the three curves represent three different level of X-rays from the source (X-ray to UV luminosity fractions $f_X$=0.0 (solid red), 0.05 (long dashed blue) and 0.1 (dotted magenta curve)), while the power law spectral index $\alpha$ is kept fixed (1.5). The three curves in second column represent three different values of $\alpha$ (0.5, 1.0 and 1.5, respectively) while $f_X$ = 0.05 is kept fixed.}
   \label{nhtktstb}
\end{center}
\end{figure}

Before proceeding to the construction of global maps, we validate our code by studying the behaviour of some basic quantities for an isolated source surrounded by a uniform IGM. Figure \ref{nhtktstb} shows the ionization, heating and $\lya$ coupling patterns around a source which is taken to be a galaxy with stellar mass $10^8\, \MSUN$ at $z = 10$. The results are shown at a time $10^7$ years after the sources began to radiate. The uniform IGM around the source is assumed to have a density equal to the mean density of the universe. The different curves in each panel represent different X-ray properties for the galaxy.

The top left panel of the figure shows the ionization fraction of hydrogen along the radial direction for different values of $f_X$. One can see that the size of the \HII region created around the galaxy is as large as $\sim 150$ pkpc.  In absence of X-rays ($f_X = 0$) the ionization front shows a sharp transition, while in presence of X-rays the transition between ionized to neutral medium is relatively smooth. The reason behind this is the presence of high energy photons in the spectrum, which have longer mean free paths and are efficient in partially ionizing the gas. Note that the increase in the X-ray luminosity does not increase the size of the fully ionized \HII region, but creates a partially ionized region of larger size beyond the \HII region. The top right panel of Fig~\ref{nhtktstb} shows the \HII fraction for different values of the X-ray spectral index $\alpha$. A larger $\alpha$ implies a steeper spectrum, which in turn implies an increase in the number of soft X-ray photons for a fixed $f_X$. As a result, the size of the partially ionized region is larger for a higher value of $\alpha$.

Panels in the second row of Figure \ref{nhtktstb} show the kinetic temperature $\TK$ pattern around the isolated source. It is clear from the left panel that the presence of X-rays affects the temperature profile quite drastically. Firstly, there is rise in $\TK$ in the ionized region when $f_X$ is increased; this is mainly due to the Compton scattering of X-ray photons with free electrons. The size of the heated region is larger than the \HII region in the presence of X-rays, with the size increasing with increasing values of $f_X$. Because of the larger mean free paths of high energy photons, $\TK$ shows a very smooth transition from the central highly heated ($\sim$ few $10^4$ K) region to far away cold region ($\sim$ few K) in the presence of X-rays, while the transition is sharp in the absence of X-rays. Even when ratio of X-ray to UV luminosity is as small as 5\%, regions as far as few hundreds of pkpc from the source have temperatures larger than the CMB temperature. With sufficient $\lya$ coupling, this heated region will show the signal in emission. An increase in $\alpha$ results in a larger number of soft X-ray photons, hence the temperature of the ionized region increases and so does the size of the heated regions as is seen from the right-hand panel of the second row.

The pattern of the spin temperature $\TS$ as calculated using equation (\ref{spinT}) is shown in the third row from top. It is clear from the plots that $\TS$ closely follows $\TK$ at distances close to the source, while it tends to follow $\TCMB$ at larger distances. This behaviour can be explained by the $d^{-2}$ decline in the $\lya$ flux $J_{\alpha}$ from the source, which makes the $\lya$ coupling stronger closer to the source and weaker at larger distances.

In order to show the differences between $\TS$ and $\TK$, we plot the function $\TS - \TK$ as a function of distance from the source in the fourth row from top. As expected, $\TS  = \TK$ close to the source because of strong $\lya$ coupling, and this relation holds for distances $\sim 300-400$ pkpc which are much larger than the \HII region. If we look at larger distances, the $\lya$ coupling weakens and $\TS$ gradually tends towards $\TCMB$. Hence one cannot work with the simple assumption that $\TS$ follows $\TK$ at early stages of reionization. Since the value of $\TS - \TK$ is essentially determined by the $\lya$ coupling, which in turn is mainly determined by $J_{\alpha}$ from the central source, we find that the curves are almost independent of the X-ray properties of the source.

The bottom row shows the 21 cm signal $\TB$ profile around the source. As is already known  \citep{thomas08, Alvarez10, Yajima13}, there are four main regions in the profile starting from the centre going outward: (i) the signal vanishes completely in the \HII region, (ii) there is an emission region just beyond the \HII region arising from X-ray heating in the partially ionized medium, (iii) the signal decreases and turns into an absorption feature because of the decrease in the value of $\TK$ and finally (iv) the absorption signal decreases and gradually vanishes as the $\lya$ coupling becomes weaker. When $f_X = 0$, the heating due to X-rays is absent. Hence the second region with emission signal is obviously absent, and the absorption signal is much stronger. As the value of $f_X$ is increased, the size of the region with emission signal increases, however the amplitude of the signal remains almost the same. The amplitude of the absorption signal in the third region decreases with increasing $f_X$. Similarly, increasing the value of $\alpha$ shows a similar effect on the emission and absorption signal.

\subsection{Global maps}
\label{global_map}

Having discussed the generation of ionization and temperature maps around an isolated galaxy, we now discuss the method for generating such maps in the full simulation box. Our treatment closely follows that of \citet{Thom09} to account for overlap between the individual patterns, though we have introduced some modifications in the method which are discussed below. 

The method starts by listing the location and mass of all the sources in the box. One then determines the radius $R_{\rm HII}$ of the \HII region, defined as the distance at which $x_{\rm HII}$ falls to $x_{\rm HII}^{\rm th} = 0.5$, for each source. Since the ionization profile shows a sharp drop from $x_{\rm HII} = 1$ to $x_{\rm HII} \sim 0.1$ for the values of $f_X$ and $\alpha$ considered in this paper, varying the threshold value in the range $0.1 <  x_{\rm HII}^{\rm th} < 1$ should not affect our results.  For a given source the average \HI number density is measured within a sphere of radius $R_0$ around the source. Initially $R_0$ is taken to have a small value. Having found the overdensity within $R_0$, we find out the radius of the ionization front $R_1$ for the source  from the 1-D catalogue as obtained in Section \ref{rt}. If $R_0$ is taken to be sufficiently small, we usually end up with $R_1 > R_0$. Since the sources form preferentially at high-density peaks, we expect the average density within the sphere to decrease with increasing radius. Hence we take $R_1$ to be the next guess for $R_0$ and iterate the process till $R_1 \approx R_0$. We then assign the radius of the \HII region to be $R_{\rm HII} = R_1$. Let us denote the radius of the \HII region and the mean hydrogen number density within the \HII region for $i^{th}$ source to be $R^{i}_{\rm HII}$ and $n^{i}_{\rm H-1D}$, respectively. 

 Having found the quantities $R^{i}_{\rm HII}$ and $n^{i}_{\rm H-1D}$, the correct ionization and heating profile around the source can be selected from the previously generated catalogue of 1-D profiles. While assigning the ionization profile within the \HII region is straightforward, one needs to be slightly careful while assigning the ionization fraction in the partially ionized regions. If one assigns the same profile corresponding to $n^{i}_{\rm H-1D}$ in the partially ionized region, as is done by \citet{Thom09}, then there will be an underestimation in the value of ionization fraction. This is due to the fact that the average density decreases with distance from the source. In order to account for this effect, we assume that the number of photons from the $i^{th}$ source at a point $\mathbf{x}$ in the partially ionized region is given by $x^{i}_{\rm HII-1D}({\mathbf x}) \times n^{i}_{\rm H-1D}$, where $x^{i}_{\rm HII-1D}({\mathbf x})$ is the ionization fraction as obtained from our catalogue of 1-D profiles. If the number density of hydrogen in that particular pixel is $n_{\rm H}({\mathbf x})$, then the ionization fraction will be modified to $x^{i}_{\rm HII-1D}({\mathbf x}) \times n^{i}_{\rm H-1D} / n_{\rm H}({\mathbf x})$. Note that we do not account for the fact that the number of recombinations will also be less for lower densities as that would involve modelling of sub-grid physics.

When the individual \HII regions overlap there will be excess photons residing at the overlapped regions which need to be accounted for. Following \citet{Thom09} we use an iterative process to estimate the total number of excess photons in the overlapping regions and distribute them among the contributing ionizing sources. The main difference between our approach and that of \citet{Thom09} arises while assigning the ionization profile in the partially ionized regions. In the presence of X-rays, these regions are larger than the \HII region and hence their overlap begins much earlier. The ionization fraction in overlapping partially ionized regions is given by
\begin{equation} 
x_{\rm HII}({\mathbf x}) =\frac{\sum_i x^{i}_{\rm HII-1D}({\mathbf x}) \times n^{i}_{\rm H-1D}}{n_{\rm H}({\mathbf x})}.
\label{overlapion}
\end{equation}

%......................................................

The heated regions too extend well beyond the \HII regions, and hence they start overlapping very early during reionization. Let a point be heated up by photons from $n$ sources. Let $\{T_1,T_2,..,T_n\}$ be the set of temperatures at that point obtained from the catalogue of 1-D profiles for these sources. In such a situation, a possible approach could be to assign the temperature in the overlap region by invoking the conservation of energy \citet{Thom09}. In our study, however, we have used a different approach which is more straightforward to implement. 

\begin{figure}
\begin{center}
\includegraphics[scale=0.565]{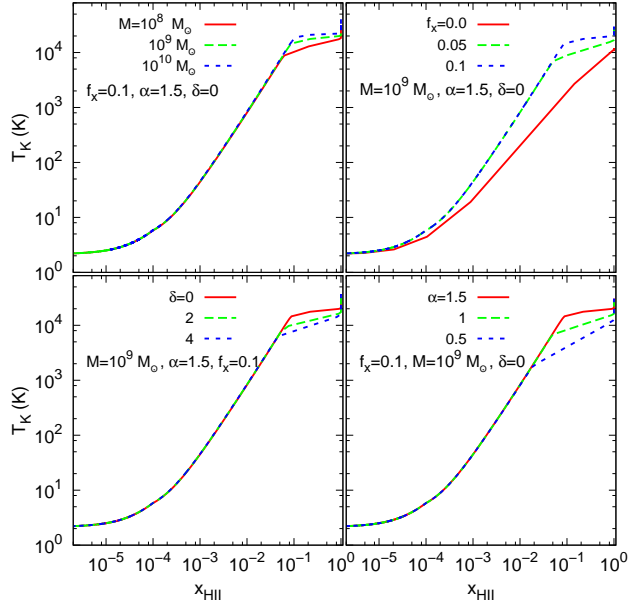}
    \caption{The kinetic temperature as a function of ionization fraction for an isolated source for different source masses (top left), X-ray to UV luminosity ratios (top right), IGM density contrasts (bottom left) and X-ray power law spectral indices (bottom right) at $z=10$ at a time 10 Myr after the source starts radiating.}
   \label{corel}
\end{center}
\end{figure}

Our approach is based on the correlation between $\TK$ and $x_{\rm HII}$ in the heated regions for isolated sources. Figure~\ref{corel} shows the plot of $\TK$ as a function of $x_{\rm HII}$ for different types of source properties. The top-left panel shows the plot for different stellar masses in the galaxy, the bottom-left panel shows the same for different values of the density contrast, the top-right is for different values of $f_X$ and the bottom-right panel shows the plots for different values of $\alpha$. Clearly a strong correlation exists between $\TK$ and $x_{\rm HII}$ for $x_{\rm HII} < 0.1$ and this correlation is completely independent of the source properties and densities for $f_X > 0$. Thus we can use this correlation for determining temperature in the overlapped regions given that we already know how to estimate the ionized fraction. This method is expected to be inaccurate for $0.1 < x_{\rm HII} < 1$, however, the fraction of such points is negligible as can be seen from the top panel of Figure \ref{nhtktstb}. In addition these points are highly heated and have strong $\lya$ coupling, thus making the signal almost independent of $\TK$. The correlation is somewhat different when $f_X = 0$. In such cases, however, we find that the temperature falls sharply beyond the \HII regions and hence the IGM can be treated as a two-phase medium characterized by the complete ionized and neutral regions. Thus we can still estimate the $\TK$ from the correlation without introducing any significant error.

 Finally, we discuss how to deal with points which receive $\lya$ radiation from more than one source. In this case, accounting for overlap is almost trivial as $J_{\alpha}$ essentially measures the number of $\lya$ photons and hence one just has to add the fluxes from different sources \citep{thomas11}.

%.......................................................

\subsection{Redshift space distortion}
\label{rsd}
Observations of the redshifted 21-cm radiation at a specific frequency can be mapped to a redshift and thus to a position along the line of sight. In absence of the peculiar velocities of \HI gas, 21 cm radiation is redshifted only due to the expansion of the universe. The redshift will be modified in the presence of the peculiar velocities of \HI gas. As gas tends to move towards overdense regions, over/underdense regions will appear more over/underdense at large scales. This changes the strength of the observed 21 cm power spectrum and makes it anisotropic \citep[see][for detailed reviews of the theory]{bharadwaj05, barkana05a, mao12}.

The method for implementing the effect of peculiar velocities on the reionization 21 cm signal during the `emission' phase (i.e, $\TS \gg \TCMB$) is reasonably well studied \citep{mellema06, mao12, Jensen13, Majumdar13}.  The situation is different in our case because at high redshift there are regions where $\lya$ coupling is not sufficiently strong and regions which may not be highly heated. We follow a method outlined in \citet{mao12} with certain modifications to account for the fluctuations in the spin temperature. We know the positions and peculiar velocities of all the dark matter particles at a certain redshift in our simulation box. Let the $i^{th}$ dark matter particle have position ($x^i,y^i,z^i$) and velocity ($v_x^i,v_y^i,v_z^i$) and have an associated hydrogen mass $M_{\rm H}^i$. We also know the neutral fraction of hydrogen and spin temperature at the grid point that contains the particle. For the scenario where $\TS \gg \TCMB$, the mass of the neutral hydrogen associated with the $i^{th}$ particle can be written as
\begin{equation}
M_{\rm H I}^i=M_{\rm H}^i x^i_{\rm H I} 
\end{equation}
where $x^i_{\rm HI}$ is the neutral fraction of hydrogen associated to the particle. This is sufficient in case the only fluctuations in $\TB$ arise from the neutral hydrogen field. If we want to account for the spin temperature fluctuations, we need to modify the above relation suitably. Since the fluctuations in $\TB$ arises from the combination $x_{\rm HI}~(1 - \TCMB/\TS)$, we define an effective \HI mass associated with the particle as 
\begin{equation}
\tilde{M}_{\rm H I}^i=M_{\rm H}^i x^i_{\rm H I}  \left(1 - \frac{\TCMB}{\TS}\right).
\end{equation}

If the line of sight is taken to be along the $x$-axis, the position of the particle $\mathbf{s}$ in redshift space coordinate will be given by,
\begin{equation}
s_x^i=x^i + \frac{v_x^i(1+z_{\rm obs})}{H(z_{\rm obs})},\\
s_y^i=y^i,\\
s_z^i=z^i
\end{equation}
where $z_{\rm obs}=(1+z_{\rm cos})(1-v_x^i/c)^{-1} - 1$ is the observed redshift and $z_{\rm cos}$ is the cosmological redshift. Once this mapping from $\mathbf{x} \to \mathbf{s}$ is established, we interpolate the \HI contributions $\tilde{M}_{\rm HI}^i$ of each particle to an uniform grid in the redshift-space. The resultant $\TB$ map will contain the effects of redshift space distortions which would correspond to the observed signal.

%\\\\\\\\\\\\\\\\\\\\\\\\\\\\\\\\\\\\\\\\\\\\\\\\\\\\\\\\\\\\\\\\\\\\\\\\\\\\\\\\\\\\\\\\\\\\\\\\\\\\\\\\\\\\\\\\\\\\\\\\\\\\\\\\\\\\\\\\\\\\\

\section{Results }
\label{res}

We present the results of our analysis of the 21 cm signal in this section. As mentioned in section \ref{21cm}, the signal depends on the neutral hydrogen fraction, kinetic temperature, the $\lya$ coupling and the line of sight velocity of the neutral gas. The model we have used accounts for all these effects. However, it is important to note that there exist studies which simply assume $\TS \gg \TCMB$, which is obtained when the medium is highly heated ($\TK \gg \TCMB$) and $\lya$ coupling is very strong ($\TS = \TK$). In such cases, the effect of fluctuations in $\TS$ on the signal can be ignored. Similarly, many studies ignore the effect of redshift-space distortion.  While presenting our results using the model where all the effects are accounted for, we will also look into the effects of not accounting for the fluctuations in $\TS$.
\begin{figure}
\begin{center}
\includegraphics[scale=0.7]{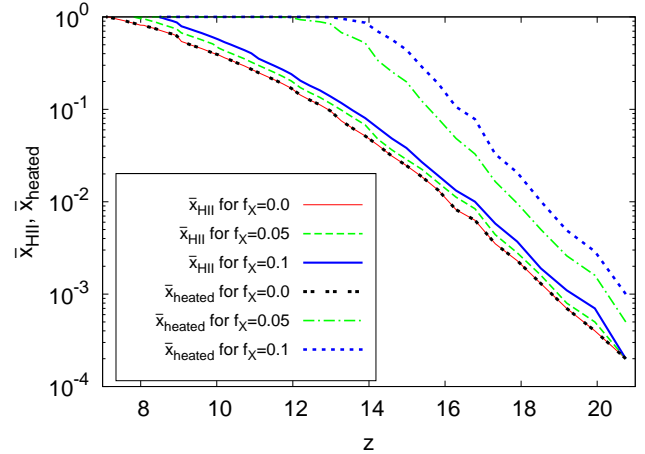}
    \caption{ Evolution of volume weighted ionization fraction of hydrogen and heated fractions of the IGM for different X-ray luminosities of the source. The thin solid red, dashed green and thick solid blue curves represent the evolution of $x_{\rm HII}$ for $f_X=$ 0, 0.05 and 0.1 respectively. The value of $\alpha$ is taken to be 1.5. The fractional volume occupied by heated regions (defined as those with $\TK > \TCMB$) are represented by the other three curves.  These models are for the scenario where reionization is driven by haloes identified using spherical overdensity halo-finder in the simulation box.}
   \label{ionfrc}
\end{center}
\end{figure}

\subsection{ Reionization driven by only high mass sources}
\label{large_halo}
 In this section, we present the nature of signal if reionization is driven solely by high mass sources.  These sources are identified using spherical overdensity method from the $N$-body simulation. The minimum mass for this study is $3.89 \times 10^9\, h^{-1}\, \MSUN$. We will discuss the effect of small sources on the signal later in a separate section.

\subsubsection{Global ionization and heating history}
\label{res:global_hist}

Before carrying out any analysis, we need to fix the value of the star-forming efficiency $f_\star$. As mentioned earlier, the value of $f_\star$ would determine the global reionization history. Hence, we fix its value by demanding that it matches the electron scattering optical depth $\tau$ as observed by the CMB observations. We find that, in the absence of X-rays ($f_X = 0$), the choice $f_\star = 0.03 $ gives $\tau = 0.08$ which is consistent with the observed constraints \citep{Hinshaw09}. Adding X-ray photons to the luminosity increases the value of $\tau$, however, the effect is quite small. For example, using $f_X = 0.1$ with $\alpha = 1.5$ increases $\tau$ to 0.089 which still is consistent with the CMB constraints. Hence, we fix the value $f_\star = 0.03$ and concentrate on studying the effects of changing the values of $f_X$ and $\alpha$. The evolution of the ionized hydrogen fraction $x_{\rm HII}$ is shown in Figure \ref{ionfrc}. Most of the reionization process occurs between redshifts 16 and 8 during which the ionization fraction grows from 0.01 to 1. Since our simulation boxes are not equipped for treating small mass haloes, the reionization occurs faster than what would be allowed by quasar spectra constraints at $z \approx 6$ \citep{Gunn65, fan03, fan2006, Malhotra06, Choudhury06a, Bolton2007}. However, since the main purpose of this work is to study the effects of peculiar velocities on the 21 cm signal, we did not concern ourselves too much on reproducing the reionization constraints. As mentioned earlier, we have also checked whether ignoring the small mass haloes at early stages has any effect on our conclusions, which we will discuss later in Section \ref{small_halo}.

We can see from Figure \ref{ionfrc} that increasing the value of $f_X$ causes reionization to occur earlier, though the effect is not that drastic (at least for $f_X < 0.1$, the highest value of $f_X$ considered in this paper). However, the heating pattern is very sensitive to the value of $f_X$. As can be seen from the figure, the volume fraction in which the kinetic temperature is above the CMB temperature reaches unity only at $z \approx 8$ (at the same time when reionization is complete) for $f_X = 0$, while the same happens much earlier, at $z \approx 12 (14)$ for $f_X = 0.05 (0.1)$.

We have considered a number of models in this paper which depending on how the various effects have been accounted for. These models are summarised in Table \ref{tab1}. Unless otherwise specified, the parameters related to the X-ray sources are fixed to be $f_X = 0.05$ and $\alpha = 1.5$. For model A, we have assumed the IGM to be uniformly heated and the $\lya$ coupling to be highly efficient, thus making the signal independent of $\TS$. Model B accounts for the fact that the IGM may not be uniformly heated, i.e., the pattern of $\TK$ is calculated self-consistently, however the $\lya$ coupling is still taken to be highly efficient, thus making $\TS =  \TK$. In model C, the $\lya$ coupling too is calculated self-consistently accounting for the inhomogeneities in the $\lya$ flux $J_{\alpha}$. These models are similar to those considered by, e.g., \citet{baek09}.

\begin{table}
\centering
\small
\tabcolsep 3pt
\renewcommand\arraystretch{1.2}
   \begin{tabular}{c c c}
\hline
\hline
    Model   &$\lya$ coupling  & Heating  \\
\hline
\hline
    A      &Coupled        & Heated    \\
    B     &Coupled        & Self-consistent        \\
    C    &Self-consistent      & Self-consistent   \\
\hline
\end{tabular}
\caption[]{Different kinds of models considered in this paper. The terms `coupled' and `heated' represent the scenarios $x_{\alpha} \gg 1$ and  $\TK$ $\gg$ $\TCMB$ respectively.}
\label{tab1}
\end{table}

\begin{figure}
\begin{center}
\includegraphics[scale=0.7]{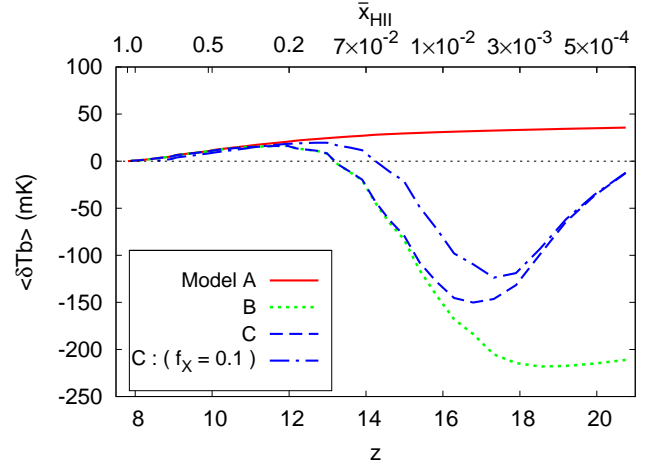}
\caption{ Evolution of volume weighted brightness temperature for model A (solid red curve), B (dotted green curve) and C (dashed blue curve) with redshift. The X-ray properties are chosen such that $f_X = 0.05$ and $\alpha = 1.5$. The dash-dotted blue curve represents model C with a higher value of $f_X = 0.1$. These models are for the scenario where reionization is driven by haloes identified using spherical overdensity halo-finder in the simulation box.}
   \label{tbev}
\end{center}
\end{figure} 

The evolution of the volume averaged brightness temperature $\TB$ is shown in Figure \ref{tbev}. For model A, the IGM is assumed to be heated and $\lya$ coupled, thus the brightness temperature is always positive and essentially traces the neutral hydrogen distribution. Once the non-uniform heating is accounted for (model B), the signal shows absorption at earlier times. This represents colder regions in the IGM where the X-ray flux may not have percolated as yet. As the fraction of regions with X-ray heating increases, the signal shows up in emission and follows model A at later stages. Since the $\lya$ coupling is assumed to be strong, $\TS = \TK$, and hence the signal is always non-zero (except for the point where $\TK = \TCMB$). When the effects of $\lya$ coupling are treated self-consistently (model C), the $\TS$ can depart from $\TK$ towards $\TCMB$ at earlier stages when the coupling is not efficient enough, and hence $\TB$ tends to vanish. As soon as the ionized fraction $x_{\rm HII} \sim 0.05$, the $\lya$ coupling seems to be sufficiently strong so that models C and B becomes identical. We also show in the figure the $\TB$ evolution when the fraction of X-rays is increased $f_X = 0.1$ in model C. Clearly, the effects of heating are visible relatively early in reionization history. It is interesting to note that all the models A, B and C are identical when $x_{\rm HII} \gtrsim 0.2$ and the assumption of $\TS = \TK \gg \TCMB$ works extremely well in these stages. It is only at very early stages that we need to account for the fluctuations in the spin temperature \citep{santos08, baek09, thomas11, Pritchard12, Mesinger2013}.

\subsubsection{Fluctuations in the brightness temperature}
\label{res:tb}

\begin{figure*}
\begin{center}
\includegraphics[scale=0.7]{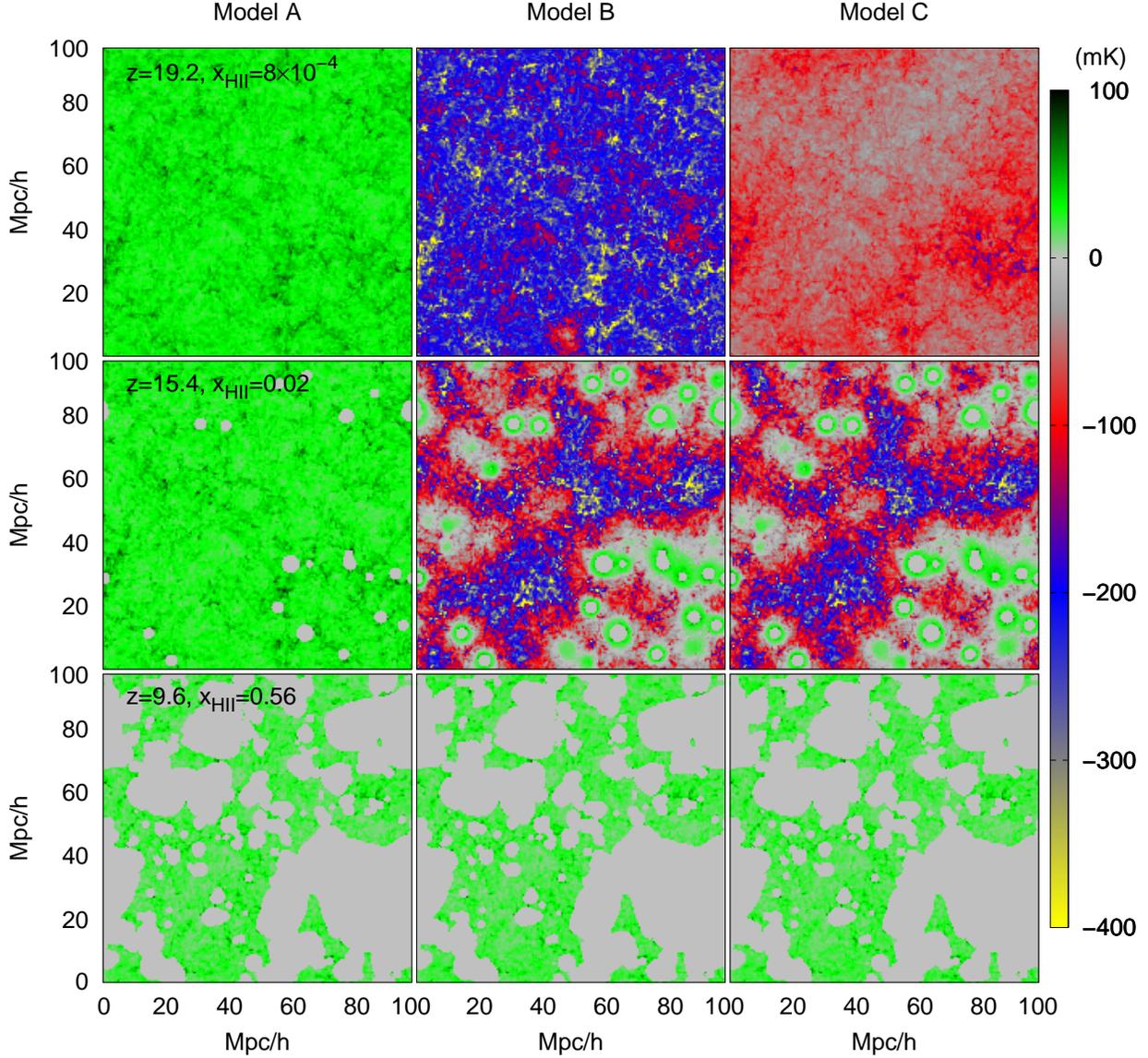}
    \caption{The brightness temperature maps for three different redshifts 19.2, 15.4 and 9.6 for model A,B and C for the reionization scenario driven by haloes identified using spherical overdensity halo-finder. Model A assumes the IGM to be $\lya$ coupled and highly heated ($\TS \gg \TCMB$). Model B assumes the IGM to be strongly $\lya$ coupled but self-consistently heated with X-ray source properties $f_X=0.05$ and $\alpha=1.5$ ($\TS \approx \TK$), while model C considers $\lya$ coupling and X-ray heating self-consistently.}
   \label{tbslic}
\end{center}
\end{figure*}

Since the main target of the radio interferometers is to measure the fluctuations in the 21 cm signal, we discuss how different effects impact the fluctuations. The maps of $\TB$ obtained from a random slice through our simulation box is shown in Figure \ref{tbslic}. The left-hand panels show the maps for model A for three different redshifts $z = 19.2, 15.4, 9.6$ respectively. Since this model corresponds to the case $\TS \approx \TK \gg \TCMB$, the brightness temperature essentially follows the neutral hydrogen distribution. In the absence of any significant sources at $z=19.2$, the IGM is neutral and the fluctuations are mainly those corresponding to underlying dark matter density field. One can see spherical ionized bubbles appearing in the slice at $z=15.4$, where the signal drops to zero. These bubbles percolate in the IGM and gives the patchy reionization map at $z = 9.6$ as is expected.

The middle panels of Figure \ref{tbslic} show the brightness temperature maps for model B where it is assumed that $\TS \approx \TK$ but the temperature $\TK$ is estimated self-consistently. At early stages $z=19.2$, the map for model B is very different from that for model A because the effect of X-ray heating is not that strong. In fact, most of the IGM shows up in absorption because it is colder than the CMB. Once the sources form, the signal shows a number of features. One can identify the spherically-shaped regions around the sources where the signal vanishes as expected. However, there is a ``ring'' of emission which corresponds to the regions which are heated by X-ray from the sources. As the effect of X-rays decreases away from the sources, the emission signal drops, changes sign and shows up in absorption. At this stage, the maps have both emission and absorption features \citep{santos08,Baek2010,thomas11,Mesinger2013}. As we come to lower redshifts $z = 9.6$, the X-ray heating is dominant all over the IGM and model B is almost identical to model A.

The right-hand panels of Figure \ref{tbslic} show the maps of $\TB$ for the model C. The main difference between this model and B is at very early stages $z = 19.2$. Since the $\lya$ coupling is calculated self-consistently using the value of $J_{\alpha}$, the map shows effects of inefficient coupling when sources are sparse. As a result, the magnitude of the absorption signal is less than that in model B. However, since one requires only a small amount of $\lya$ radiation for efficient coupling, we find that the differences between model B and C go away by $z = 15.4$.

\begin{figure*}
\begin{center}
\includegraphics[scale=0.7]{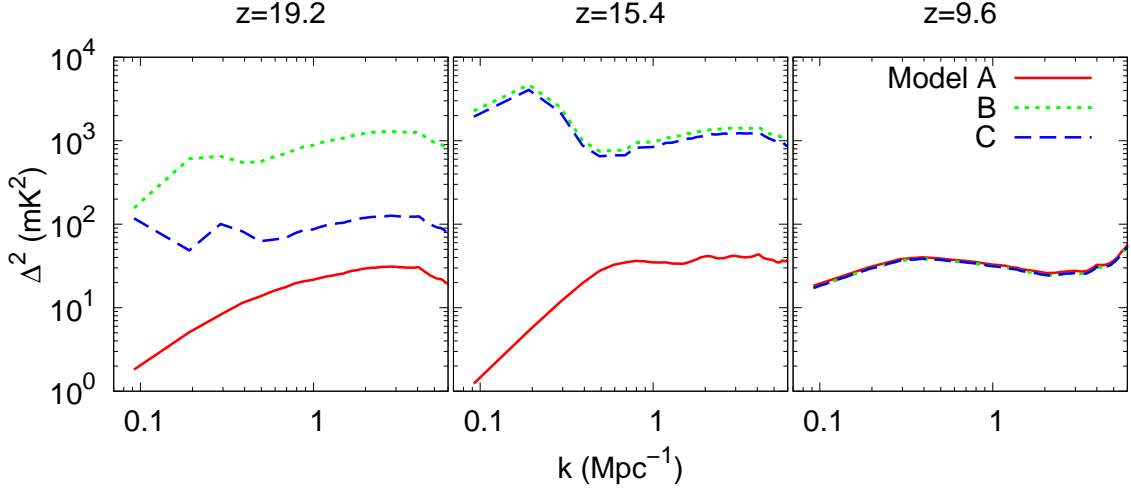} 
    \caption{ The power spectrum of brightness temperature for models A (solid red), B (dotted green) and C (dashed blue) for the scenario where reionization is driven by haloes identified using spherical overdensity method. Left to right-hand panels represent previously considered three redshifts $z =$ 19.2, 15.4 and 9.6 respectively. }
   \label{multbps}
\end{center}
\end{figure*}

The same conclusions can be drawn if we plot the power spectrum of $\TB$ for the three models, which is done in Figure \ref{multbps}. It is clear that at relatively later stages of reionization $z = 9.6$, all the models overlap with each other when the signal traces the HI distribution. At $z = 15.4$, models B and C differ from A because of the inhomogeneities in the X-ray heating which is not accounted for in A. Since there are regions showing strong absorption signals in models B and C, it leads to a stronger contrast in the maps, and hence the amplitude of the power spectrum is much larger (almost $\sim 30$ times). Interestingly, the power spectra for these models show a ``bump'' or a peak around $k_{\rm peak} \approx 0.2$ Mpc$^{-1}$. This scale corresponds to the typical sizes of regions which are heated by the sources, i.e., the sizes of the heated bubbles. In fact, one can estimate the typical radius of heated regions as $R_{\rm heat}=2.46/k_{\rm peak}$ \citep{friedrich11}, which in our case turns out to be $\sim 12$ Mpc. The amplitude of this peak is determined by the contrast in the signal between heated (i.e., emission) and colder (i.e., absorption) regions. This feature was noted in simulations of \citet{baek09} too. Since the peak has a relatively high amplitude ($\sim 5000 \, {\rm mK^2}$) it can easily be detected by an instrument like the SKA. The size of a typical heated region is an important parameter which can help in constraining the nature of X-ray sources such as the SED and the total X-ray flux.  At the earliest stages ($z = 19.2$), the effects of inhomogeneous $\lya$ coupling makes model B different from C. As is clear, because of inefficient $\lya$ coupling in regions away from sources, the brightness temperature amplitude becomes smaller in model C compared to B. This leads to a decrease in the power spectrum amplitude.

\begin{figure}
\begin{center}
\includegraphics[scale=0.7]{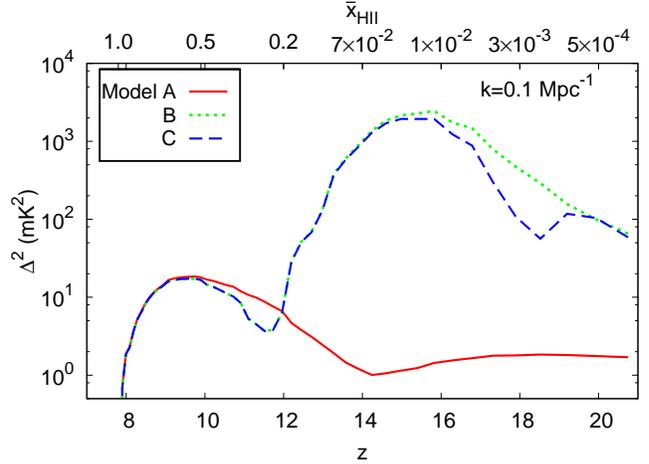}
    \caption{ Evolution of $\TB$ power spectrum at scale $k = 0.1~{\rm Mpc}^{-1}$ for model A (solid red), B (dotted green) and C (dashed blue) respectively. These results are for the scenario where reionization is driven by haloes identified using spherical overdensity method. }
   \label{psz}
\end{center}
\end{figure}

\begin{figure*}
\begin{center}
\includegraphics[scale=0.7]{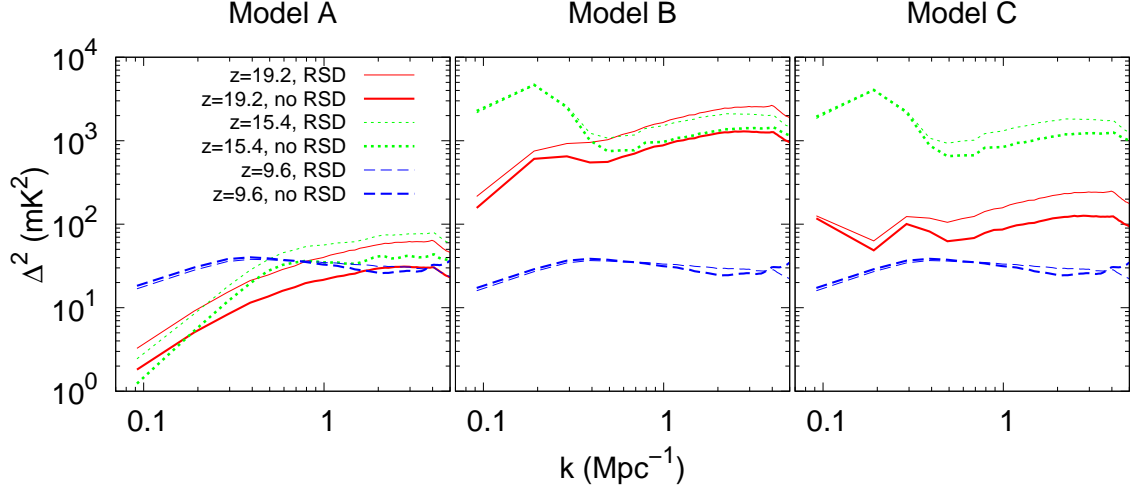}
\caption{ Effects of redshift space distortion on $\TB$ power spectra for models A, B and C where reionization is driven by haloes identified using spherical overdensity method. The thick curves represent the power spectrum without redshift space distortion at three redshifts 19.2 (solid red), 15.4 (dotted green) and 9.6 (dashed blue) respectively.  The thin curves represent the corresponding power spectrum with redshift space distortion.  }
\label{rsdts}
\end{center}
\end{figure*}

The evolution of the power spectra for different models for a typical scale $k = 0.1~{\rm Mpc}^{-1}$ accessible to first generation low-frequency telescopes is shown in Figure \ref{psz}.
For model A, the amplitude decreases with increase in ionization fraction during initial stages of reionization. However as the characteristic size of the ionized bubbles increase with time, $\Delta^2$ starts to increase. This leads to a prominent trough-like feature at $z \approx 14.2$ when the ionization fraction is $\sim 0.05$. The rise in $\Delta^2$ at $z  <14$ is halted when the bubbles start overlapping and the patchiness in the ionization fraction decreases. At this point, the amplitude is determined by $x_{\rm HII}$ and hence starts falling as reionization enters its last stages. Thus a prominent peak at $z \sim 10$  is created in the evolution curve of the power spectrum when the IGM is around 50 $\%$ ionized. 

At early stages of reionization the power spectra for models B and C are $\sim$ 100-1000 times larger than predicted by model A. This is simply due to the fact that the signal is in absorption for models B and C while for model A it is in emission. This difference reduces over time and almost vanishes when the universe is half ionized around redshift 10.

In addition to the peak arising from $x_{\rm HII}$ fluctuations, there is another peak at $z \sim 16$ for model B. This peak arises because of $\TK$ fluctuations due to inhomogeneous X-ray heating. Since the heating is not efficient at high redshifts, the signal is in absorption and hence larger than that in model A. As sources heat up the IGM, the amplitude of the signal decreases and tends towards model A. Once sufficient heating is completed in the IGM, the signal follows model A. Model C is quite similar to B except that there is another peak in the amplitude at $z \sim 19$. This peak corresponds to inhomogeneities in $\lya$ coupling which is not accounted for in the other models. Once the $\lya$ coupling becomes efficient at $z \sim 16$, the signal for model C follows B.

%RSD-------------------------------------------------------------------

\subsubsection{Effects of redshift-space distortion}
\label{res:rsd}

\begin{figure*}
\begin{center}
\includegraphics[scale=0.75]{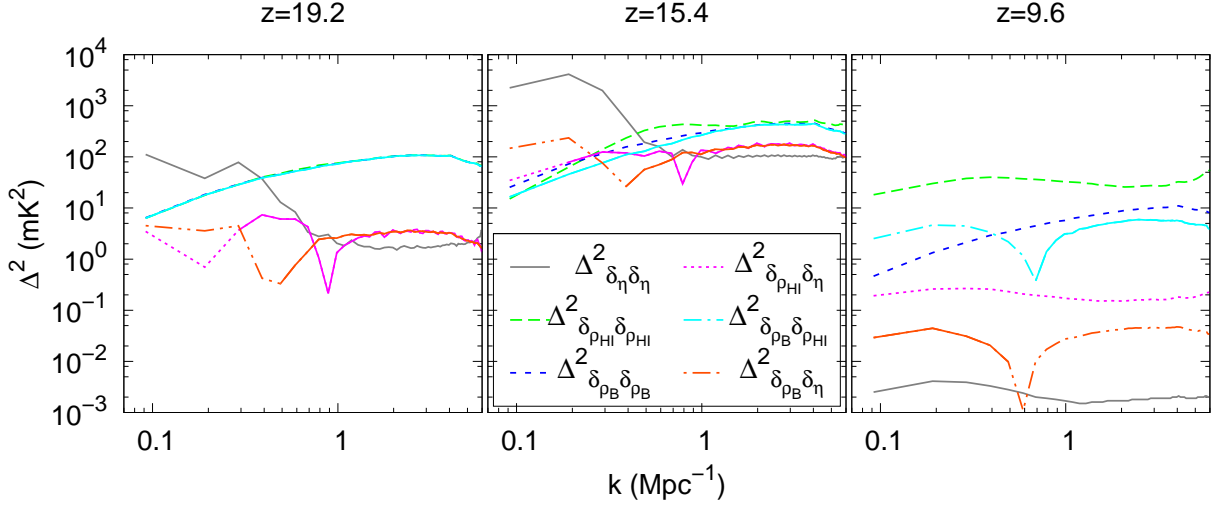}
\caption{The auto and cross-power spectra of baryonic density contrast $\delta_{\rho_{\rm B}}$, neutral hydrogen density contrast $\delta_{\rho_{\rm HI}}$ and the contrast $\delta_{\eta}$ in the term $\eta = 1 - \TCMB/\TS$  at three different redshifts. The solid parts in the cross-power spectra curves represent positive values, while the other parts represent negative values. The reionization is assumed to be driven by haloes identified using spherical overdensity method.}
\label{pskmuall}
\end{center}
\end{figure*}

As is well known, the effects of including the peculiar velocities in the brightness temperature calculation are two-fold: one is to introduce anisotropies in the signal, and the other is to modify the amplitude of the spherically averaged power spectrum. In this work, we will mostly study the second effect, i.e.,  the change in the amplitude because of redshift-space effects. We will also comment on the anisotropies wherever appropriate.

Before proceeding to presenting our results, let us briefly discuss a quasi-linear model \citep{mao12} for studying the effects of redshift space distortion (RSD). In order to see this, let us write equation (\ref{brightnessT}) as 
\begin{equation}
\TB (z,\mathbf{x}) = \widehat{\TB}(z) \bar{\eta}(z)[1+\delta_{\rho_{\rm HI}}(z,\mathbf{x})] [1+\delta_{\eta}(z,\mathbf{x})],
\end{equation}
where 
\begin{equation}
\widehat{\TB}(z) = 27 ~ \bar{x}_{\rm HI}(z) \left(\frac{\OmegaB h^2}{0.023}\right) \left(\frac{0.15}{\Omegam h^2}\frac{1+z}{10}\right)^{1/2}\,\rm{mK},
\end{equation}
is the average brightness temperature at $z$ and 
\begin{equation}
\eta(z,\mathbf{x}) = 1 - \frac{\TCMB(z)}{\TS(z,\mathbf{x})}.
\end{equation} 
The average value of $\eta$ is denoted as $\bar{\eta}$ while the corresponding contrast is 
given by $\delta_{\eta} = \eta/\bar{\eta} - 1$.
The power spectrum in the redshift space can be decomposed as 
\begin{equation}
P^s(\mathbf{k}) = P_0(k) + P_2(k)~\mu^2 + P_4(k)~\mu^4,
\end{equation}
where $\mu = k_{\parallel}/|\mathbf{k}|$, $k_{\parallel}$ being the component of $\mathbf{k}$ along the line of sight. In the quasi-linear approximation, the different components appearing in the above equation are given in terms of the real-space power spectra as 
\begin{eqnarray}
\label{quasi_li}
P_0(k) \!\!\!\! &=&\!\!\!\! \left(\widehat{\TB} \bar{\eta}\right)^2 \left[
P_{\delta_{\rho_{\rm HI}} \delta_{\rho_{\rm HI}}}(k) + P_{\delta_{\eta} \delta_{\eta}}(k) + 2 P_{\delta_{\rho_{\rm HI}} \delta_{\eta}}(k)
\right],\nonumber\\
P_2(k) \!\!\!\!&=&\!\!\!\! 2 \left(\widehat{\TB} \bar{\eta}\right)^2 \left[
P_{\delta_{\rho_{\rm B}} \delta_{\rho_{\rm HI}}}(k) +  P_{\delta_{\rho_{\rm B}} \delta_{\eta}}(k)
\right],\nonumber\\
P_4(k) \!\!\!\!&=&\!\!\!\! \left(\widehat{\TB} \bar{\eta}\right)^2 
P_{\delta_{\rho_{\rm B}} \delta_{\rho_{\rm B}}}(k),
\end{eqnarray}
where $\delta_x = x/\bar{x} - 1$ is the contrast of the quantity $x$. The quantities $\rho_{\rm B}$ and $\rho_{\rm HI}$ represent the density of baryons and neutral hydrogen respectively. The quantities of the form
$P_{\delta_x \delta_x}$ denote the real-space auto-power spectrum of the quantity $x$, while  $P_{\delta_x \delta_y}$
are the real-space cross-power spectrum between fields $x$ and $y$.
 The spherically-averaged power spectrum in redshift-space is simply given by
\begin{equation}
P^s_{\rm ave}(k) = P_0(k) + \frac{1}{3} P_2(k) + \frac{1}{5} P_4(k).
\label{psave}
\end{equation}

We should mention here that we have \emph{not} used this quasi-linear model to calculate any of our results, nor have we verified its validity for our work. The model is introduced so as to help understand the various properties of the redshift-space power spectrum in a simple way.

We now present our results incorporating the effects of line of sight peculiar velocities on the $\TB$ fluctuations. The resulting power spectra for the three models are shown in Figure \ref{rsdts}. The left-hand panel shows the results for model A for three different redshifts and for the cases with and without the redshift space distortion. At higher redshifts, the effect of including the redshift space effects is to increase the amplitude of the power spectra, without affecting the shape significantly. For example, the increase in amplitude is about a factor of 1.87 at $z = 19.2$ and $15.4$ at scales probed by our simulation when the redshift space distortion is accounted for. This is expected from the quasi-linear model when there are no fluctuations in $\TS$ , i.e., $\delta_{\eta} = 0$ and the HI density follows the baryonic density field $\delta_{\rho_{\rm HI}} = \delta_{\rho_{\rm B}}$. The effect decreases as the neutral hydrogen fraction decreases further and we find that the effect is negligible for model A at $z = 9.6$. Since the difference between real-space and redshift-space power spectra for model A is essentially determined by the cross term $P_{\delta_{\rho_{\rm B}} \delta_{\rho_{\rm HI}}}(k)$, it is obvious that the correlation between the underlying baryonic field and the HI field vanishes when the ionization fronts propagate into the low-density regions at later stages of reionization.

The situation is clearly different for models B and C at higher redshifts. We find that the redshift space power spectra are higher than the real-space ones by a factor of $\sim 1.87$ at smaller scales $k \gtrsim 0.4$ Mpc$^{-1}$ at $z =$ 19.2 and 15.4 implying that the fluctuations in $\TS$ are negligible at these scales. At larger scales however, the $\TS$-fluctuations dominate the brightness-temperature power spectrum, and in addition, these fluctuations are not correlated with the baryonic density field. This makes the effect of redshift-space distortion negligible at scales $k \lesssim 0.4$ Mpc$^{-1}$. The models B and C are identical to A at $z = 9.6$, and hence the effect of redshift space is not visible.

\begin{figure}
\begin{center}
\includegraphics[scale=0.7]{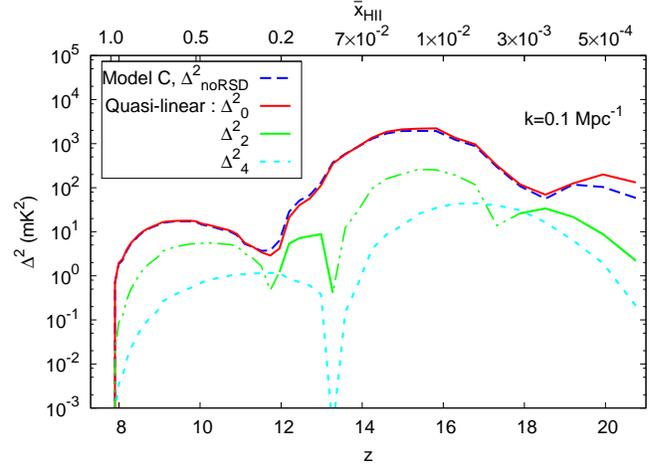}
\caption{ Evolution of different components of redshift space power spectrum (in linear theory) with redshift in model C for the reionization scenario driven by haloes identified by spherical overdensity halo-finder. The solid part in the green curve represent positive value, whereas the dotted dashed part represent negative value.}
\label{psz_mu_large}
\end{center}
\end{figure}

In order to bring more clarity to these arguments, we plot the various terms which contribute to the redshift-space power spectrum in Figure \ref{pskmuall}. We show the plots only for model C. At $z = 19.2$ and $15.4$, the power spectra will be dominated by the terms $P_{\delta_{\rho_{\rm B}} \delta_{\rho_{\rm B}}}$, $P_{\delta_{\rho_{\rm B}} \delta_{\rho_{\rm HI}}}$  and $P_{\delta_{\rho_{\rm HI}} \delta_{\rho_{\rm HI}}}$ at smaller scales, all of which are almost identical because of very little ionization. At larger scales, the fluctuations are dominated by the $P_{\delta_{\eta} \delta_{\eta}}$ term. Clearly the spin temperature fluctuations $\delta_{\eta}$ are not correlated with the other fields as is shown by the smallness of the terms $P_{\delta_{\rho_{\rm HI}} \delta_{\eta}}$ and $P_{\delta_{\rho_{\rm B}} \delta_{\eta}}$, hence we find that the difference between redshift-space and real-space power spectra are negligible. The fluctuations in $\eta$ decrease as the IGM is heated, however, because the ionization fronts have propagated into low-density regions by then, the effects of redshift-space distortions are negligible.  This can also be seen in Figure \ref{psz_mu_large}, where we have plotted the amplitude of the different components of $P^s({\bf k})$ as defined in equation \ref{psave}. It is clear that the second and fourth order anisotropic terms, due to RSD, are quite small compared to the $\mu$-independent term throughout the reionization history at large scales. Hence overall we find that the effects of redshift-space distortion are negligible at large scales $k \lesssim 0.4$ Mpc$^{-1}$, though the reasons differ over the history of reionization. At smaller scales however, the redshift space power spectrum is $\sim 2$ times higher in amplitude that the real space counterpart at early stages of reionization, mainly because the signal is dominated by the HI fluctuations at these scales. These fluctuations, in turn, are strongly correlated with the density field because of formation of sources at preferentially high density peaks.

\begin{figure}
\begin{center}
\includegraphics[scale=0.7]{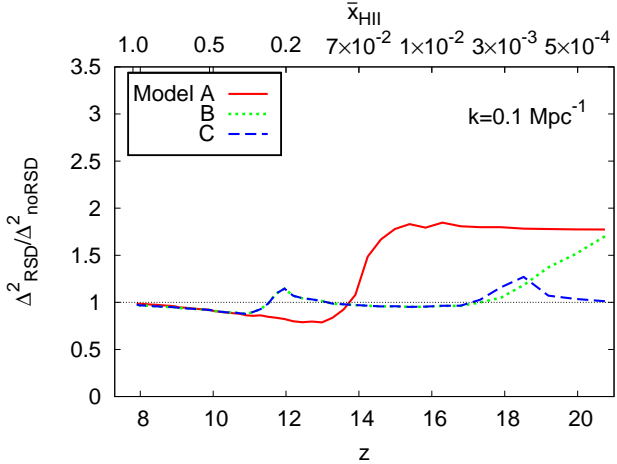}
\caption{ Ratio of power spectra with and with-out redshift space distortion as a function of redshift for models A, B and C at a scale $k = 0.1$ Mpc$^{-1}$. The plot is for the reionization scenario driven by haloes identified by spherical overdensity halo-finder.}
\label{pszratio_large}
\end{center}
\end{figure}

 The evolution of the ratio of the redshift space and real space power spectra for a typical scale $k = 0.1~{\rm Mpc}^{-1}$ accessible to first generation low-frequency telescopes is shown in Figure \ref{pszratio_large}. One obvious conclusion is that the effects of redshift-space distortion are only visible for model A and that too at early stages of reionization $\bar{x}_{\rm HII} \lesssim 0.1$. Unfortunately, this model is not appropriate at early stages of reionization because it does not account for fluctuations in $\TS$. Once these fluctuations are taken into account, the power at large scales is dominated by $\delta_{\eta}$ contribution which does not correlate with the underlying density field. Hence the effect of peculiar velocities is negligible at large scales throughout the reionization history.

 We should mention here that the results obtained using our model A are somewhat in disagreement with results obtained using radiative transfer simulations \citep{mao12,Jensen13}. For example, the maximum value of the ratio of redshift and real space power spectra is $\sim 1.9$ in our model A, while the same rises to $\sim 5$ in the case of \citet{mao12}. The main reason is that there are no small mass sources (say with halo mass $\sim 10^8\, \MSUN$) in our simulation box. As a result the ionization field is not strongly coupled to the density field. As a result, the trough in the evolution of the real space power spectrum around $\bar{x}_{\rm HII} \sim 0.2$ at large scales for model A (see Figure \ref{psz}) is not sufficiently deep \citep[compared to, e.g.,][]{mao12,Jensen13}. We note that the trough, which is very prominent in earlier works, is the main reason for the RSD power spectrum getting enhanced by a factor of ~5 for $\bar{x}_{\rm HII} \sim 0.2$ on large scales. To address this issue,  we have studied the contribution from small mass sources using a sub-grid model which is discussed in Section \ref{small_halo}.

It has been suggested that, for scenarios similar to our model A, the anisotropies in the power spectrum produced by the redshift space distortion effect is detectable in 2000 hrs of observations with LOFAR \citep{Jensen13}. It was also suggested that such observations could tell us whether reionization occurred inside-out or outside-in \citep{Jensen13, Majumdar13}. However, we find that the amplitude of the isotropic component $P_0(k)$ of the power spectrum is much larger ($\sim 10$ times, see Figure \ref{psz_mu_large}) than the anisotropic terms $P_2(k)$ and $P_4(k)$ at large scales ($k \lesssim 0.4$ Mpc$^{-1}$)  in a scenario where the spin temperature fluctuations $\delta_{\eta}$ dominate the power spectrum. Although the inclusion of $\TS$ fluctuations does not affect the amplitude of the anisotropic terms, it still may have interesting consequences for the detectability of the anisotropic signal. For example, it has been shown that the cosmic variance/sample variance is an important factor in extracting anisotropic term from the total power spectrum \citep{shapiro13}. The errors arising from cosmic variance would be proportional to $P_0(k)$ in our case, which can affect the detectability of the anisotropic components. Though extraction of anisotropy in the $\mu$-decomposition formalism could be difficult because of the sample variance and the possibility of leaking one $\mu$-term into other \citep{shapiro13, Jensen13}, there may be other ways to extract the anisotropy such as using an orthogonal basis like the Legendre polynomial formalism \citep{Majumdar13}. It could be inetersting to explore the possibility for the case where the fluctuations in the spin temperature dominate the HI 21 cm signal. It could also be interesting to explore the possibility of detecting this anisotropy in intermediate scales ($k \sim 1$ Mpc$^{-1}$) where the effects of redshift space distortion are much more significant.

\begin{figure}
\begin{center}
\includegraphics[scale=0.7]{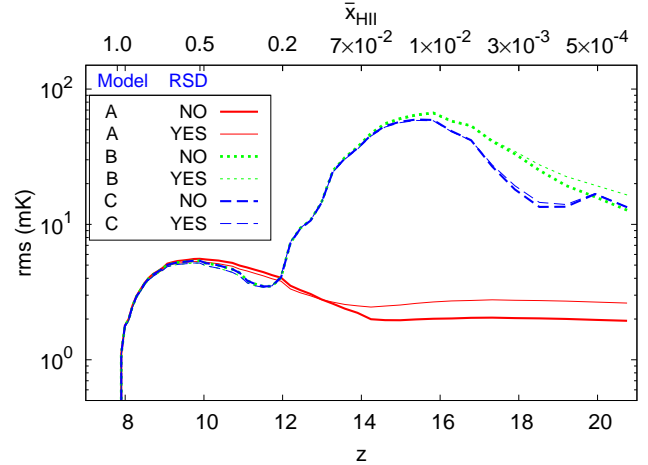}
\caption{Evolution of the rms of brightness temperature fluctuations (smoothed over scales of 15 cMpc) as a function of redshift for models A, B and C with and without the effects of redshift space distortion. The plot is for the reionization scenario driven by haloes identified by spherical overdensity halo-finder.}
\label{rms}
\end{center}
\end{figure}

The root mean square ( rms ) of HI brightness temperature fluctuations $\TB$ is another major statistical quantity that will also be targeted by the first generation instruments \citep{mellema06, jelic08, bittner11, patil2014}. It is defined as,
\begin{equation}
{\rm rms}(\TB) =\sqrt{ \frac{1}{N}\sum^{N}_{i=1} (\delta T_{b}^i -\overline{ \delta T_{b}})^2 },
\end{equation}
where $N$ is the number of $\TB$ data cube pixels after smoothing over a scale of 15 cMpc and $\overline{ \delta T_{b}}$ is the average value of $\TB$. The plot of rms deviation as a function of $z$ is shown in Figure \ref{rms}. The behaviour of this quantity, as expected, is very similar to the evolution of $\Delta^2(k)$ at large scales (Figure \ref{psz}). It shows three peaks at different locations corresponding to different physical processes discussed earlier. Recently, it was proposed that a function having a single peak (e.g.,  $\propto (z/z_R)^{\beta} \left( 1+\tanh \left[(z-z_R)/\Delta z] \right) \right)$, where $\beta$ is the power law index, $z_R$ is the location and $\Delta z$ is the width of the peak) can be fitted to the measured rms vs. $z$ data to reliably extract information about the redshift of reionization $\sim z_R$ and the duration of reionization $\sim\Delta z$ \citep{patil2014}. The conclusion was based on an reionization model similar to A where the rms vs. $z$ curve has just one peak around $x_{\rm HII}\sim 0.5$. However in Figure \ref{rms}, we find that there could be two additional peaks at large scales, due to inhomogeneous Ly$\alpha$ coupling and heating respectively.  The presence of these two additional peaks makes the interpretation of the measured rms vs. $z$ slightly more complicated. In particular, the peak corresponding to the inhomogeneous heating can have a very large amplitude and can affect the detection of the low-redshift peak in a noisy data. Fitting a single peak function over the entire range of redshifts could thus possibly result in misinterpretation of the observations. One should thus either restrict the analysis within lower redshifts (e.g., $8<z<11$ for model C) or choose a function having three peaks. One should also keep in mind that the influence of the spin temperature fluctuations in the rms (and also in the power spectrum) remains for a while even after the entire IGM is heated above the CMB temperature.

As expected, we find that the effects of redshift space on the rms are negligible when $\TS$-fluctuations are accounted for in the model (model C). This implies that the modelling of the 21 cm signal at large scales is possible without accounting for peculiar velocities of the gas, however, it is critical that one accounts for the fluctuations in $\TS$.

\subsubsection{Effect of different  X-ray source properties on 21 cm power spectrum}
\label{res:X_rsd}

It is clear from the previous section that the redshift-space effects are negligible when the X-ray heating and $\lya$ coupling are accounted for self-consistently. It is necessary to check whether this result is independent of the parameters related to the X-ray emissivities, which we do in this section. We will essentially consider model C  where all the physical effects are accounted for self-consistently, and vary the parameters $f_X$ and $\alpha$ to see the effects on the real and redshift-space power spectrum. In particular, we would concentrate on the evolution of the power spectrum amplitude at a scale $k = 0.1~{\rm Mpc}^{-1}$.

\begin{figure}
\begin{center}
\includegraphics[scale=0.7]{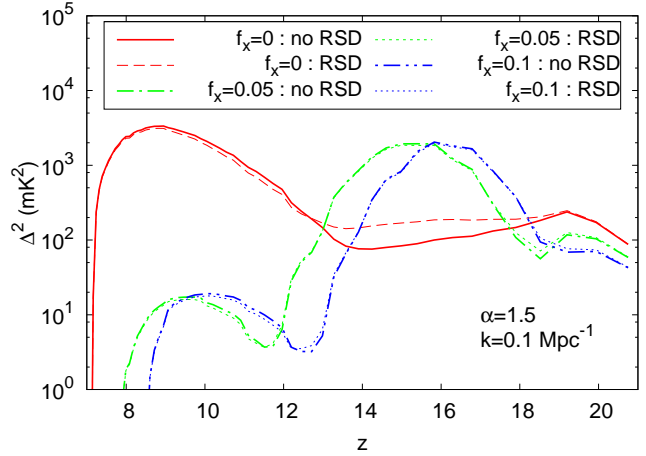}
\caption{Evolution of $\delta T_{b}$ power spectra at $k = 0.1$ Mpc$^{-1}$ for three different values of $f_X$ with and without the effects of redshift space distortion. These models have only haloes which could be identified using the spherical overdensity halo-finder.}
\label{fxpsx2}
\end{center}
\end{figure}

Let us first consider the effect of varying $f_X$ and take three representative values $f_X = 0, 0.05, 0.1$. The corresponding power spectra are shown in Figure \ref{fxpsx2}.   When $f_X = 0$, the neutral hydrogen will contain almost no photons of higher energies and hence there would be almost no heating. As a result, the neutral regions will be visible in absorption while the ionized regions will show no signal. This also means there will be no fluctuations in $\TK$ once the $\lya$ coupling is complete. This is clear from the figure which shows that the case with $f_X = 0$ contains the initial peak at $z \sim 20$ arising from inhomogeneities in the  $\lya$ coupling. However, as expected the model does not have the subsequent peak corresponding to the $\TK$ fluctuations. The amplitude again peaks because of the increase in bubble sizes and then decreases as reionization is completed. The amplitude at these last stages, interestingly, is much larger compared to the X-ray heated models which is because of the fact that the IGM is cold and shows strong absorption signal. This model is essentially same as the model A discussed above but the mean brightness temperature $\overline{\TB}$ being replaced by the mean kinetic temperature of the IGM. When $f_X > 0$, the signal shows the prominent peak corresponding to the $\TK$ fluctuations. It is expected that the amount of heating and the size of the heated regions would increase as the X-ray intensity increases. This suggests that the IGM will be heated at an earlier redshift when $f_X$ in increased which is clearly seen from Figure \ref{fxpsx2}. In fact, the peak arising from $\TK$ fluctuations appear earlier when the value of $f_X$ is increased. Hence the position of the peak as a function of redshift (or frequency of radio observations) can be used for probing the level of total X-ray background. This is consistent with earlier results \citep{christian13}.

As far as the effect of redshift-space distortion is concerned, we find that there is some difference between the redshift and real space amplitudes at $13 < z < 18$ when $f_X = 0$. This is arising because there are no fluctuations in the spin temperature at these epochs and hence the model is similar to model A. However, the moment when $f_X > 0$, the effect of redshift-space distortion vanishes and becomes independent of the actual value of $f_X$.

\begin{figure}
\begin{center}
\includegraphics[scale=0.7]{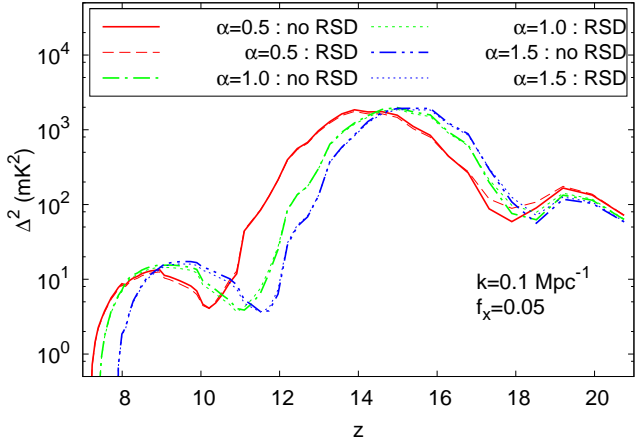}
\caption{Evolution of $\TB$ power spectra with redshift at scale $k = 0.1$ Mpc$^{-1}$ for three different values of the spectral index $\alpha$ of the miniquasar SED. The results are shown for cases with and without the effects of redshift space distortion. The value of $f_X$ is kept fixed to 0.05. The plot is for the reionization scenario driven by haloes identified by spherical overdensity halo-finder.}
\label{alphapsx2}
\end{center}
\end{figure}

The conclusions remain similar when we vary the value of the spectral index $\alpha$. The effect of considering different values of $\alpha$ is shown in Fig. \ref{alphapsx2}. We find that the peak related to $\TK$ fluctuations appear earlier when $\alpha$ is increased. This is because for the same value of $f_X$, increasing $\alpha$ means redistributing photons to relatively lower energies. Since lower energy photons are easier to be absorbed by HI, the amount of ionization and heating is more. The most interesting aspect to check is the effect of redshift space distortions, and we find that the effect is negligible independent of the value of $\alpha$.

\begin{figure}
\begin{center}
\includegraphics[scale=0.7]{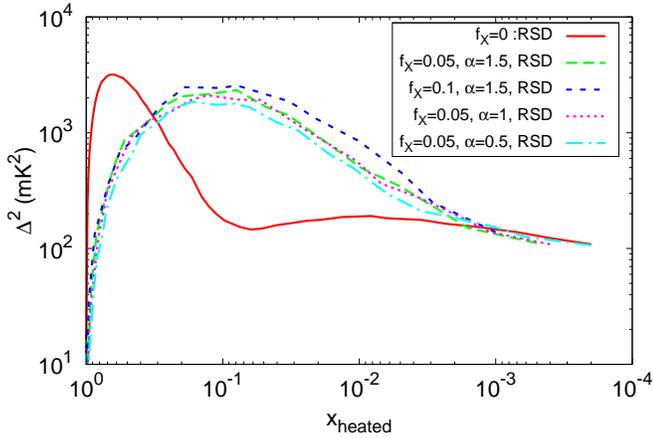}
\caption{Evolution of the $\TB$ power spectra at $k = 0.1$ Mpc$^{-1}$ as a function of heated volume fraction in the IGM. Regions with $\TK > \TCMB$ are referred to as heated regions. The plot is for the reionization scenario driven by haloes identified by spherical overdensity halo-finder.}
\label{psfheat}
\end{center}
\end{figure}

\begin{figure}
\begin{center}
\includegraphics[scale=0.7]{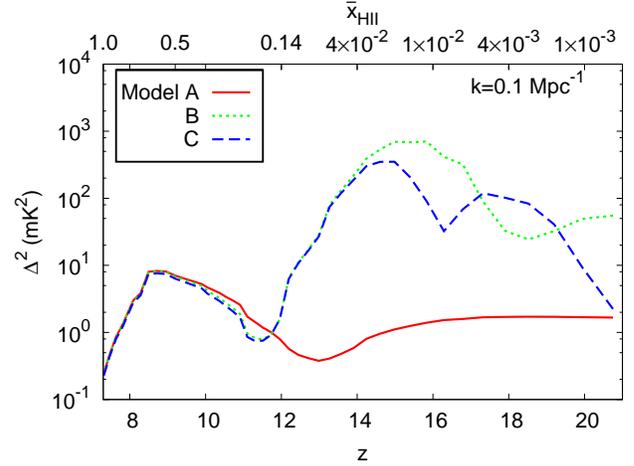}
\caption{The brightness temperature power spectra as a function of redshift for models A, B and C at a scale $k = 0.1$ Mpc$^{-1}$ for the scenario where small mass haloes are taken into account using a subgrid prescription.}
\label{rsdz_small}
\end{center}
\end{figure}

It is clear that our results on redshift space distortion is independent of the exact X-ray background chosen, as long as it is non-zero. The only effect of varying $f_X$ and $\alpha$ is to change the location of the peak in the power spectrum amplitude arising from $\TK$-fluctuations. In Fig. \ref{psfheat}, we show the evolution of the power spectrum amplitude as a function of the heated fraction $x_{\rm heated}$ for different values of $f_X$ and $\alpha$. We should mention that the quantity $x_{\rm heated}$ is defined as the fraction of points which have $\TK > \TCMB$. It is interesting that for $f_X > 0$, the location of the peak is independent of the level of X-ray background or the spectral index of the X-ray sources and appears when $x_{\rm heated} = 0.1$. It would be interesting to discuss the possibility of detecting this peak using the next generation of radio telescopes as that would clearly establish the signature of fluctuations in the IGM heating. Additionally, the position of the peak will tell us about the redshift when $10\%$ of the IGM volume was heated above the CMB temperature.

\begin{figure*}
\begin{center}
\includegraphics[scale=0.7]{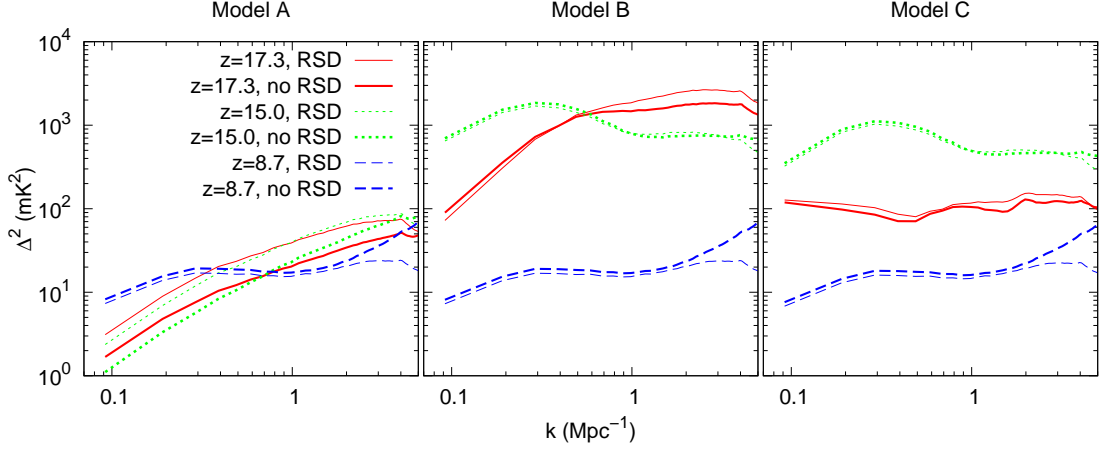}
\caption{The brightness temperature power spectrum as a function of scale at three different redshift for model A, B and C in a scenario which accounts for small mass haloes through a subgrid prescription.}
\label{rsdk_small}
\end{center}
\end{figure*}

%%%%
\subsection{ Effect of small mass haloes}
\label{small_halo}

 In this section, we describe the effects of small mass haloes on the 21 cm signal, particularly at large scales. The initial stages of reionization are expected to be driven by sources in haloes as small as $\sim 10^8\, \MSUN$. One difficulty with the models we have considered so far is that they do not contain haloes smaller than $3.89 \times 10^9\, h^{-1}\, \MSUN$ (set by the particle mass of the simulation and the halo-finding algorithm), and hence the ionization field is not sufficiently correlated with the underlying density field. In order to address this issue, we use a sub-grid prescription to identify haloes with $M_{\rm halo} < 3.89 \times 10^9\, h^{-1} \MSUN$ inside the $100\, h^{-1}$ Mpc box. The details of the prescription is discussed in Appendix \ref{appendix}. The results in this section are obtained using the parameters $f_X = 0.05$ and $\alpha = 1.5$. We have varied the parameter $f_{\star}$ to obtain a reionization history identical to that in \ref{res:global_hist}.\footnote{We have not included any effects of radiative feedback on the small mass haloes.}

The evolution of the brightness temperature power spectra for the three models A, B and C at a scale $k = 0.1$ Mpc$^{-1}$ is shown in Figure \ref{rsdz_small}. One can see that the broad features are similar to that obtained for the scenario with only high mass haloes. During the initial stages of reionization, the ionization field is highly correlated with the density field and hence the trough in the evolution curve for model A is deeper than previously obtained in Figure \ref{psz}.  In presence of a large number of small mass sources, the patchiness in the ionization field is decreased as expected. The overall amplitude of the fluctuations in the brightness temperature is also lower at large scales compared to our fiducial case. As a result the amplitude of the heating peak is lower than previously obtained in model B and C. For example, the heating peak in the evolution of the power spectrum is $\sim$ 5 times smaller than that in the previous scenario (reionization driven by only high mass sources).  We also see a significant difference in the power spectrum when plotted as a function of scale shown in Figure \ref{rsdk_small}. First of all, peakiness of the "bump" like feature at large scale noted earlier is somewhat diminished. This is because the fluctuations are less when small sources are taken into account. As the number of small mass sources are very large in this scenario, the characteristic heated region is also smaller than what was obtained previously at high redshift.  As a result the scale corresponding to the ``bump'' shifts, in general, towards the smaller scales  when smaller sources are taken into account at high redshift. Like in the previous scenario, the heating peak (in the evolution curve of the power spectrum) occurs when $\sim$ 10 $\%$ gas (by volume) is heated above $\TCMB$.

\begin{figure}
\begin{center}
\includegraphics[scale=0.7]{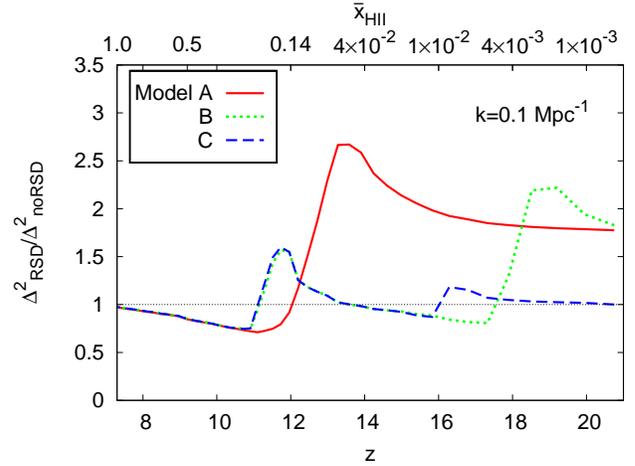}
\caption{Ratio of redshift and real space power spectrum as a function of redshift at large scale for model A, B and C (small mass sources included using a subgrid prescription).}
\label{rsdz_ratio}
\end{center}
\end{figure}

For model A, the effect of redshift space distortion is similar to the results of previous scenario (as shown in Figure \ref{rsdk_small}). In order to understand the effect of RSD, we have plotted the ratio of the redshift space and real space power spectra at $k = 0.1$ Mpc$^{-1}$ as a function of $z$ in Figure \ref{rsdz_ratio} for the three models A, B and C. In this case the ratio increases initially, which was not found in previous scenario (see Figure \ref{pszratio_large}). This ratio increases to $ \sim 3$  when the universe is $ \sim  20\%$ ionized by mass for model A, which is similar to the results of previous studies like \citet{mao12, Majumdar13, Jensen13}. This increment of the ratio depends on the level of coupling of the ionization field to the density field. We also checked with a source model similar to \citet{Jensen13} and we found the ratio increases even more than 3 for model A.

\begin{figure}
\begin{center}
\includegraphics[scale=0.7]{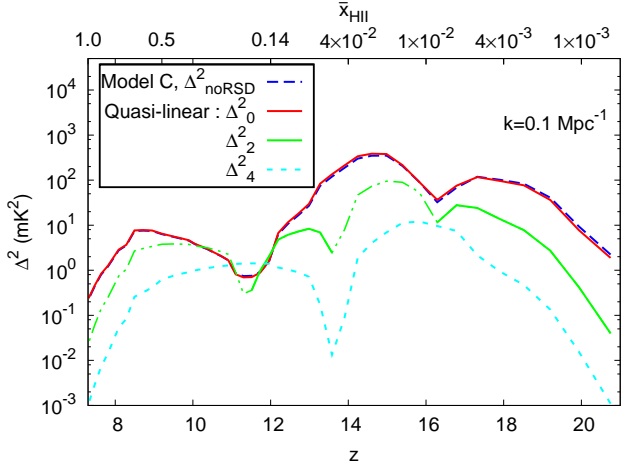}
\caption{Evolution of different components of redshift space power spectrum (in linear theory) with redshift in model C for a scenario where small mass sources are included using a subgrid prescription. Solid part in the green curve represent positive value, whereas the dot-dashed part represent negative value.}
\label{psz_mu_small}
\end{center}
\end{figure}

For models B and C, the effect of redshift space distortion is similar at large scales, as shown in Figure \ref{rsdk_small}. The effect of redshift space distortion is smaller even in small scales $k \gtrsim 0.4$ Mpc$^{-1}$, which was not seen previously. The reason behind this is the fact that, with large number of small sources taken into account, the fluctuations in brightness temperature at small scales too is dominated by $\TS$-fluctuations. In another words, the power spectrum at redshift $\sim 15$ is dominated by the term $P_{\delta_{\eta} \delta_{\eta}}$ (see equation \ref{quasi_li}) even at small scales $k \sim 3$ Mpc$^{-1}$.

The ratio of the redshift space and real space power spectrum for model C, when plotted as a function of redshift, shows negligible effects for most of the time as can be seen from Figure \ref{rsdz_ratio}. The effect of RSD is prominent only when the power spectrum have minima in its evolution, but this effect is for short period. As shown in Figure \ref{psz_mu_small}, the anisotropies in the power spectrum are quite smaller (except for near the minima or dip in the evolution) than the real space power spectrum at large scales and thus the effect of redshift space distortion is not able to boost the power spectrum significantly at large scale for model C.

%%%%%

\subsection{Effect of box size on the 21 cm power spectrum}
\label{small_box}

As mentioned earlier, our fiducial simulation volume does not contain haloes smaller than $3.89 \times 10^9\, h^{-1}\, \MSUN$ which are believed to be the main drivers of reionization at early stages. We have tried to overcome this difficulty by including small haloes using a sub-grid prescription in Section \ref{small_halo}, however such prescriptions may not be completely accurate. Resolving such small haloes using any halo-finder requires simulations of very high dynamic range. In order to address this issue, we run an additional dark matter simulation of size 30 $h^{-1}$ cMpc with $768^3$ particles. The mass resolution achieved in this case is $5.254 \times 10^6\, h^{-1}\, \MSUN$ , thus giving a minimum halo mass of $1.05 \times 10^8\, h^{-1}\, \MSUN$. This box is helpful in probing the effects of small mass haloes at early stages and also study the effects of box size and resolution. One should, however, note that this box is too small for studying the large scales which are expected to be accessible to the first generation radio telescopes. In this section, we compare the results from the 100 and $30\, h^{-1}$ cMpc boxes, where in both cases the haloes are identified using the spherical overdensity algorithm.

\begin{figure}
\begin{center}
\includegraphics[scale=0.7]{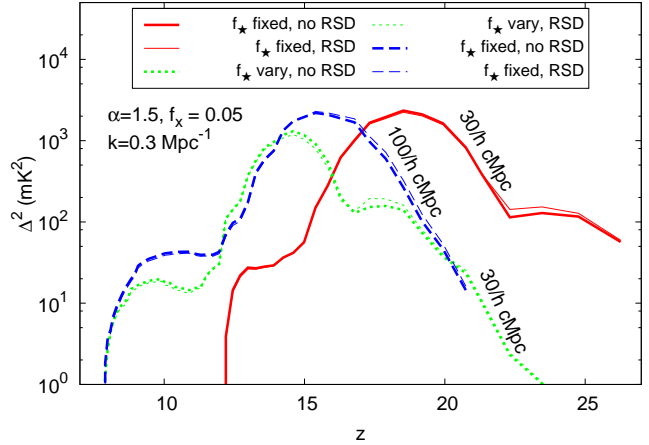}
\caption{Comparison of the evolution of $\delta T_{b}$ power spectra at $k = 0.3$ Mpc$^{-1}$ for two simulations with $30\, h^{-1}$ cMpc and $100\, h^{-1}$ cMpc boxes. The results are shown for cases with and without the effects of redshift space distortion. }
\label{30mpc}
\end{center}
\end{figure}

 Figure \ref{30mpc} show the effects of including small haloes in the analysis. The dashed blue curve represents the evolution of $\Delta^{2}$ at a scale $k = 0.3~{\rm Mpc}^{-1}$ for the fiducial model in $100\, h^{-1}$ cMpc box, while the solid red curve represents the same for the $30\, h^{-1}$ cMpc box with $f_\star$ kept same as that for  $100\, h^{-1}$ cMpc box\footnote{We should mention that we have plotted the power spectra at scale $k=0.3~{\rm Mpc}^{-1}$ in Figure \ref{30mpc} rather than at $k = 0.1~{\rm Mpc}^{-1}$ as we had done in earlier sections. This is because larger scales are not accessible in the smaller $30\, h^{-1}$ cMpc box.}. For the same value of the efficiency parameter $f_\star$, the presence of small haloes drive the reionization faster than what is found with the $100\, h^{-1}$ cMpc box.  However, the basic features, i.e., the three peaks in the power spectrum remain almost similar in the smaller box. What is interesting to note is that the effect of redshift space distortion at these scales is negligible even when the small mass haloes are taken into account.

We also study the case where $f_\star$ is varied for the small box so as to recover the ionization history identical to what we studied using the $100\, h^{-1}$ cMpc box. The dotted green curve in Figure \ref{30mpc} represents the results for such scenario. We find that the plots with and without the redshift space distortion fall on top of each other. Thus the conclusion that the peculiar velocities do not affect the large scale power remains unchanged.

%\\\\\\\\\\\\\\\\\\\\\\\\\\\\\\\\\\\\\\\\\\\\\\\\\\\\\\\\\\\\\\\\\\\\\\\\\\\\\\\\\\\\\\\\\\\\\\\\\\\\\\\\\\\\\\\\\\\\\\\\\\\\\

\section{Summary and discussion}
\label{conc}

 The main aim of this paper is to investigate the effects of peculiar velocities of the neutral gas and spin temperature fluctuations in the IGM  on the spherically averaged power spectrum  of 21 cm brightness temperature during the early stages of reionization, i.e, cosmic dawn. We have developed a code, based on the methods of \citet{thomas08,Thom09}, to generate self-consistent brightness temperature maps. Our method consists of the following main steps: (i) We generate the dark matter density and velocity field using a $N$-body code {\sc cubep}$^3${\sc m} \citep{Harnois12}. The same simulation data are used for identifying locations and masses of collapsed haloes. Since the main purpose of this work is to study the effects of peculiar velocities, it is important to model the velocity fields and the correlation between density and velocity fields as accurately as possible, which is only achieved in $N$-body simulations. (ii) The sources of reionization are modelled assuming each source has a stellar component and a mini-quasar like component for producing photons. The stellar component is modelled using a population synthesis code {\sc pegase2} \citep{Fioc97}, while the other component was assumed to have a power-law spectrum. (iii) The radiative transfer was implemented using a one-dimensional code for isolated sources, and we have accounted for overlaps in ionized and heated regions of different sources appropriately.

We have validated our code by comparing with various existing results and found that all the features expected in the fluctuation power spectrum at early stages of reionization are nicely reproduced in our calculations \citep[see e.g.][]{Pritchard07,santos08,Baek2010,mesinger2011,christian13}. In particular, we have studied in some detail the effects of inhomogeneities in the gas temperature and the $\lya$ coupling. Each of these effects produce distinct peak-like features in the large scale power spectrum when plotted as a function of redshift. The peak which appears the latest in the history corresponds to the fluctuations in the HI field, which is targeted by the present generation of radio telescopes. The second peak corresponds to fluctuations in heating and occurs when $\sim 10$\% of the volume is heated above the CMB temperature. The third peak, which occurs the earliest in reionization history, arises because of inhomogeneities in the Ly$\alpha$ coupling.

Since the spin temperature fluctuations during the early stages of reionization introduce two additional peaks in the rms vs. redshift plot, fitting a single peak function over the entire redshift range like the one proposed earlier \citep{patil2014} could possibly result in misinterpretation of the data. One can, however, restrict the analysis within lower redshift (e.g, $8<z<11$ for model C) or choose a function with three peaks. One should also be careful that the influence of the spin temperature fluctuations in the rms (and also in the power spectrum at large scales) remains for a while even when the entire IGM is heated above the CMB temperature.

 We have considered two different scenarios: (i) when reionization is driven by only high mass sources and (ii) when small mass sources are included as well. In scenario (i), the heated bubble sizes are large and as there are only rare high mass sources, fluctuations in the brightness temperature are larger at large scale compared to small scale. This causes a ``bump'' like feature in the power spectrum when plotted as a function of $k$, in a situation when the spin temperature is coupled to the kinetic temperature but the gas is heated inhomogeneously. The ``bump'' corresponds to typical size of the heated bubbles. Thus, this peak can be used to constrain the sizes of heated bubbles which has further implications in constraining properties of X-ray sources. In presence of large number of small sources, the fluctuations in $\delta T_b$ is much larger than scenario (i) at small scale and thus the ``bump'' is smoothed out to some extent and shifts towards small scales.

Once we incorporate the effects of peculiar velocities in our model and compare with the case when such effects are absent, we find that at large scales, i.e., $k \lesssim 0.4$ Mpc$^{-1}$, the effects on the spherically averaged power spectra are negligible throughout reionization history for scenario (i). It is not difficult to understand the reasons for this: the effect of peculiar velocities is substantial only when the 21 cm fluctuations are correlated with the underlying baryonic density field. At late stages of reionization ($x_{\rm HII} \gtrsim 0.2$), the 21 cm fluctuations are dominated by the fluctuations in HI field. The ionization fronts at these epochs have percolated in the low-density cosmological density field and the HI field is not correlated with the density field any more. Hence one finds no effect of the peculiar velocities on the 21 cm signal. On the other hand, during early stages of reionization ($x_{\rm HII} \lesssim 0.2$), the fluctuations at large scales are dominated by the $\TS$-fluctuations, which too are only very mildly correlated with the density field. Interestingly, if the $\TS$-fluctuations are not accounted for in the model, the 21 cm fluctuations are dominated by HI fluctuations even during early stages. In the inside-out models of reionization, the HI fluctuations are highly correlated with the density field and thus one seems to find relatively stronger effects of peculiar velocities on the 21 cm fluctuations \citep[see e.g.][]{Jensen13, Majumdar13,Majumdar2014}.
 In fact for reionization scenario with only high mass sources, at smaller scales $k  \gtrsim 0.4$ Mpc$^{-1}$, the 21 cm fluctuations even at the early stages of reionization are dominated by patchiness in the HI field, and it is not surprising that the power spectrum in the redshift space are enhanced in the amplitude compared to the real space counterpart. In presence of small mass sources, 21 cm fluctuation is dominated by $\TS$-fluctuation and thus the effect of RSD is not significant even at small scales. It was suggested that, for a reionization scenario where the spin temperature is much higher than the CMB temperature, the redshift space distortion effect which makes the power spectrum anisotropic is detectable with 2000 hrs of observations using LOFAR. It was also suggested that such observations could tell us whether reionization occurred inside-out or outside-in \citep{Jensen13, Majumdar13}. However, since the isotropic component of the power spectrum dominates over the anisotropic terms in scenarios where spin temperature fluctuations are important, the detectability of the redshift space distortion effect in the power spectrum needs to be reassessed.

We have checked our conclusions for different X-ray properties of the sources (i.e., the amount of X-rays produced compared to the UV and the steepness of the X-ray spectra), and also on the resolution of the simulation box. The conclusions seem to be quite robust in this respect. It thus implies that the 21 cm fluctuations would be quite isotropic at large scales throughout the reionization epoch, while one expects some departures from isotropy at relatively smaller scale $k \gtrsim 0.4$ Mpc$^{-1}$ if reionization is solely driven by high mass sources.

In future, it would be interesting to study the anisotropies in the 21 cm power spectrum arising from peculiar velocities at early stages of reionization. While we expect that the detectability of anisotropies at large scales could be somewhat challenging because of cosmic variance, it may be easier to detect the anisotropic signal at smaller scales. The exact nature of this signal depends on the lowest mass source at that epoch. It would be interesting to explore the possibilities of constraining early source properties using this feature. Also, one should keep in mind that we have concentrated only on peculiar velocity effects, while there could be other sources of anisotropy in the signal. A prominent effect is that of the evolution of ionization history i.e., the light cone effect \citep{Datta2012b, zawada14, plante13, Datta2014}, which we plan to address in future. The only X-ray sources considered in this paper are mini-quasar like. It would be interesting to extend our analyses to other types of X-ray sources (e.g., high-mass X-ray binaries) and verify if the conclusions remain unchanged.

\section*{Acknowledgments}
RG would like to thank Aritra Basu, Narendra Nath Patra and Prasun Dutta for valuable discussions and suggestions regarding numerical computations. KKD thanks the Department of Science \& Technology (DST), India for the research grant SR/FTP/PS-119/2012 under the Fast Track Scheme for Young Scientist. We would like to thank the anonymous referee and Suman Majumdar for their constructive comments on the paper which helped us to improve the paper.

\bibliography{mainbbl}
\bibliographystyle{mn2e}

\appendix
\section{Small mass haloes using sub-grid model}
\label{appendix}
The dark matter haloes, identified from the simulation of size $100\, h^{-1}$ cMpc using spherical overdensity method, have masses $ > 3. 89\, \times 10^9\, h^{-1}\, \MSUN$. In principle one should consider the contribution from small mass haloes during the initial stages of reionization, as they dominate the halo abundance and can drive the reionization very differently. In this appendix, we intend to investigate the effects of small mass haloes on the signal without compromising on the box size. In order to introduce the small mass haloes (say, between $\sim 10^8\, \MSUN$ and $\sim 10^9\, \MSUN $) in our box, we have used a sub-grid recipe, which is described in this appendix.

In this appendix, we describe the method used to obtain small mass haloes (i.e., smaller than that allowed by the dark matter particle mass of the simulation) using a sub-grid recipe. This is mainly done using the extended Press-Schechter model of \citet{Bond1991} and hybrid prescription of \citet{barkana2004}. The steps are as follows. 

\begin{itemize}

\item[(1)] First, the non-linear overdensity at each grid point is calculated from the mass density distribution, obtained from the $N$-body simulation, 
\begin{equation}
\delta_{\rm NL}(z,{\bf x}) = \rho(z,{\bf x})/\bar\rho(z) - 1,
\end{equation}
where $\rho(z,{\bf x})$ is the mass density at comoving position ${\bf x}$ at redshift $z$ and $\bar\rho(z)$ is the mean mass density at that redshift. 

\item[(2)] If the comoving volume and non-linear overdensity of a cell are $V_{\rm cell}$ and $\delta_{\rm NL}$ respectively, then the Lagrangian cell size $R_{\rm cell}$ corresponding to that cell is given by the relation
\begin{equation} 
\frac{4\pi}{3}R^{3}_{\rm cell}\times  \bar\rho(z) =  (1+\delta_{\rm NL})V_{\rm cell}\times \bar\rho(z).
\end{equation}

\item[(3)] The linearly extrapolated overdensity $\delta_{L}$, corresponding to $\delta_{\rm NL}$, is calculated by solving the following parametric equations \citep[e.g.,][]{ahn2014}
\begin{equation} 
\delta_{\rm NL}=\left(\frac{10\delta_{L}}{3(1-\cos\theta)}\right)^{3}-1,\,\,\,\delta_{L}=\frac{3\times6^{2/3}}{20}\left(\theta-\sin\theta\right)^{2/3},
\label{Li_Nli_pos}
\end{equation}
when $\delta_{\rm NL}(z,{\bf x})>0$, and
\begin{equation}
\delta_{\rm NL}=\left(\frac{10\delta_{L}}{3(\cosh\theta-1)}\right)^{3}-1,\,\,\,\delta_{L}=\frac{3\times6^{2/3}}{20}\left(\sinh\theta-\theta\right)^{2/3},
\label{Li-Nli-neg}
\end{equation}
when $\delta_{\rm NL}(z,{\bf x})<0$.

\item[(4)] According to the extended Press-Schechter model \citep{Bond1991}, the collapse fraction (i.e. the fraction of mass in collapsed objects) at ${\bf x}$ is given by,
\begin{equation}
f^{\rm PS}_{\rm coll}(z, {\bf x}; M_{\rm min}) = {\rm erfc} \left(\frac{\delta_c(z) - \delta_{L}(z, {\bf x})} {\sqrt{2 (\sigma^{2}_{R_{\rm min}} - \sigma^{2}_{R_{\rm cell}})}} \right),
\end{equation}
where $\sigma^{2}_{R_{\rm cell}}$ is the variance of the mass $M_{\rm cell}$, enclosed in a sphere of comoving radius $R_{\rm cell}$ and $\sigma^{2}_{R_{\rm min}}$ is variance to the minimum halo mass $M_{\rm min}$ that can host a galaxy. The parameter $\delta_c(z)$ is the critical overdensity for halo collapse at redshift $z$. The minimum mass in our study is taken to be \citep[e.g.,][]{Barkana2001}
\begin{equation}
M_{\rm min} = 3 \times 10^9 (1+z)^{-1.5} \MSUN.
\end{equation}
This implies a threshold mass of $\sim 10^8\, \MSUN$ at $z \sim 10$, appropriate for studying haloes which can cool via atomic hydrogen transitions.

\item[(5)] While $f^{\rm PS}_{\rm coll}(M_{\rm min})$ can be worked out analytically in a straightforward manner, the mass function of \citet{Sheth1999} provides a much better match with simulations. Thus we follow  hybrid prescription of \citet{barkana2004} and normalize the collapse fraction to match with the simulation result:
\begin{equation}
f_{\rm coll}(z, {\bf x}; M_{\rm min}) = \frac{\bar f^{\rm ST}(z)}{\bar f^{\rm PS}(z)}{\rm erfc} \left(\frac{\delta_c(z) - \delta_{L}(z,\mathbf{x})} {\sqrt{2 (\sigma^{2}_{R_{\rm min}} - \sigma^{2}_{R_{\rm cell}})}} \right),
\end{equation}
where $\bar f^{\rm PS}$ and $\bar f^{\rm ST}$ are the mean collapse fraction according to Press-Schechter and Sheth-Tormen prescriptions respectively.

\item[(6)] As we have the collapse fraction at each cell inside the simulation box, the total mass within haloes with mass above $M_{\rm min}$ in each cell can be obtained from
\begin{equation}
M_{\rm halo}(z,{\bf x}) =  f_{\rm coll}(z,{\bf x}; M_{\rm min}) \times  (1+\delta_{NL}(z,{\bf x}))V_{\rm cell}\times \bar\rho(z).
\end{equation}
We prepare a new halo list, where small mass haloes (smaller than the smallest halo identified in the box, i.e., $3.89 \times 10^9\, h^{-1}\, \MSUN	$) are obtained from this sub-grid prescription and haloes heavier than $3.89 \times 10^9\, h^{-1}\, \MSUN$ are obtained directly from the simulations using spherical overdensity method.

\end{itemize}

%%%%

%\begin{appendix}
%\input{appendix}
%\end{appendix}
\label{lastpage}
\end{document}